\documentclass[%
reprint,
nofootinbib,
amsmath,amssymb,
prd,
]{revtex4-2}
\usepackage{aas_macros}
\usepackage{graphicx}
\usepackage{dcolumn}
\usepackage{bm}
\usepackage{xcolor}
\usepackage{hyperref}
\hypersetup{
  colorlinks   = true, 
  urlcolor     = blue, 
  linkcolor    = blue, 
  citecolor   = red 
}
\usepackage{tabularx}
\usepackage{lipsum}

\newcommand{\universemachine}[0]{\texttt{UniverseMachine}}

\newcommand{\vlos}[0]{v_{\rm LOS}}
\newcommand{\rlos}[0]{r_{\rm LOS}}
\newcommand{\vpeclos}[0]{v_{\rm pec,LOS}}
\renewcommand{\vr}[0]{{v_{\rm r}}}
\newcommand{\vt}[0]{{v_{\rm t}}}
\newcommand{\xijsu}[0]{\xi_{\rm JSU}}
\newcommand{\vrpk}[0]{v_{\rm r,peak}}
\newcommand{\svr}[0]{\sigma^2_{\vr}}

\newcommand{\Mpc}{{\rm Mpc}}

\begin{document}

\preprint{APS/123-QED}

\title{Cluster Infall for Mass Calibration in the Stage-IV Era}

\author{Connor Sweeney}
    \affiliation{Department of Physics, University of Arizona, Tucson, AZ 85721, USA} 
 \email{cosweeney@arizona.edu}
\author{Eduardo Rozo}%
    \affiliation{Department of Physics, University of Arizona, Tucson, AZ 85721, USA}

\date{\today}
\begin{abstract}

The outskirts of galaxy clusters present a promising avenue for constraining cluster masses in a way that is robust to the impact of baryonic physics.  We assess the accuracy to which the cluster infall regions can be used for cluster mass calibration. Building on previous work, we parameterize the velocity distribution $P(v_{\rm r},v_{\rm tan}|r,M)$ of dark matter halos on scales $r\geq 5\ h^{-1}\ \Mpc$ as the product of the marginalized distribution $P(\vr|r,M)$ and the conditional distribution $P(v_{\rm tan}|\vr,r,M)$, calibrating the radial and mass dependence of these distributions in numerical simulations. We then project our model along the line-of-sight to obtain accurate predictions for the distributions of line-of-sight velocities at a given projected radius and cluster mass $P(\vlos|R,M)$, which we can observe with spectroscopic survey data.  With  our model, we forecast that spectra from the Dark Energy Spectroscopic Instrument (DESI) can constrain cluster masses with sub-percent level precision, comparable to that of Stage IV weak lensing surveys.  

\end{abstract}

\maketitle

\section{Introduction}
 
As the most massive gravitationally bound structures in the Universe, galaxy clusters are natural probes of cosmology. In particular, the abundance of clusters as a function of mass is a powerful cosmological probe (e.g., \cite{salcedo_etal25}; see \cite{miyatake2025cosmologygalaxyclusters} for a recent review). However, exploiting this sensitivity is complicated by nonlinear processes within clusters, including feedback from active galactic nuclei and other baryonic effects. This has motivated studies of cluster dynamics at radii $r \gtrsim 2\ h^{-1}\,\mathrm{Mpc}$, which are relatively insensitive to such physics while remaining sensitive to cluster potentials and cosmic expansion. These studies have shown that the dynamics in these regions enable competitive constraints on redshift-space distortions \cite{zuweinberg13, 10.1093/mnras/staa2249}, cluster masses \cite{10.1093/mnras/stz2227}, the cosmic expansion rate \cite{10.1093/mnras/stad601}, and even modified gravity \citep{lam_etal12,zu_etal14}.

The study of the dynamical phase-space in \cite{10.1093/mnras/stad601} investigates the use of the stacked line of sight (LOS) velocity dispersions of galaxy clusters as a standard ruler, finding competitive percent-level constraints on the expansion rate. These velocity dispersions as a function of cluster-centric projected radius exhibit a characteristic `kink' associated with the extent of orbiting galaxies (see their Fig. 3). As both the amplitude of the dispersion profile and the size of cluster, as indicated by the extent of orbiting galaxies, depend on cluster mass, one can use the profile as a distance measure. This relies on a model for the distribution of projected LOS velocity distributions $P(v_{\mathrm{LOS}}|R, M)$ parameterized in terms of the so-called `edge radius' indicated by orbiting galaxies' maximum radial extent. Their approach, informed by previous work \cite{Tomooka} \& \cite{zuweinberg13}, is to model $P(v_{\mathrm{LOS}}|R, M)$ directly as a sum of contributions from three dynamically distinct galaxy populations: orbiting, infalling, and `far'. 

Here, we investigate the properties of the same LOS velocity distributions at large scales, beyond the extent of the cluster halos. Our goal is to develop an accurate model for the line-of-sight velocity distributions, and to determine whether these distributions have the potential to be useful for precision cluster mass calibration in the Stage IV era. If successful, this work will establish the foundation for a research program centered on this idea.  

Our approach diverges from that in \cite{10.1093/mnras/stad601} in a number of ways: first, we make no distinction between infalling and `far' galaxies, finding `far' galaxies to simply be the part of the infalling population sufficiently influenced by expansion. Additionally, since it is trivial to find the distribution of the LOS velocities from those of the \textit{peculiar} LOS velocities (the non-Hubble flow component), we are interested in a model that accounts for the distributions of both. This approach faces a number of challenges, however. First, the increasing magnitude of the mean infall velocity near the edge radius leads to a characteristic broadening of the peculiar LOS velocity distribution which is difficult to parameterize as a function of projected radius alone. Similarly, it is difficult to find a comprehensible parameterization as a function of both projected radius and LOS separation. For these reasons, we decide to follow the approach of previous work that is able to neatly parameterize such distributions as functions of the three-dimensional radial separation, namely \cite{zuweinberg13}, \cite{10.1093/mnras/stz2227}, and more recently \cite{2024MNRAS.533.4081R}. These studies model the joint distribution of the radial and tangential peculiar velocities which have components along the LOS at a given cluster-centric radius. The joint distribution is further composed as a mixture of two components that contribute at difference scales, corresponding to a `virialized' or `splashback' component and an infalling component. 

In this work, we focus constructing a more accurate and parametrically compact description of the infalling component, with the goal of enabling kinematic mass calibration that is robust against baryonic effects.  We are heavily informed by previous work in following this methodology; however, we attempt to make model simplifications where possible in order to reduce the large number of parameters  needed to describe the joint distributions at the percent level as functions of radius and mass in \cite{2024MNRAS.533.4081R} \citep[and therefore][]{zuweinberg13}.  We then use the resulting velocity probability distributions to determine the accuracy with which pairwise-velocity data in the cluster-infall regions can constrain clusters masses with DESI, updating the forecasts from \citep{10.1093/mnras/stz2227}.

\par We structure the remainder of this paper as follows: in Section ~\ref{sec:sims}, we describe the methodology used for our study, including the simulation data and the selection cuts used to construct our synthetic galaxy catalog. We outline our updated model for the joint distributions at a given mass in Section ~\ref{sec:VD}, and then extend it to account for mass dependence. We then detail the model used for the density profiles of galaxies needed for computing the projected LOS velocity distributions from the joint distributions in Section ~\ref{sec:cf}. We compute the projected LOS velocity distributions in Section ~\ref{sec:LOSVD}, and discuss their potential to competitively constrain cluster masses in Section ~\ref{sec:calib}. Lastly, we conclude in Section ~\ref{sec:summary}.  

\section{Simulation Data}
\label{sec:sims}

We use the Multi-Dark Planck 2 (MDPL2) simulation, consisting of $3840^3$ dark matter particles of mass $1.51\times10^{9} M_{\odot}h^{-1}$ in a box with side length 1 $h^{-1}$ Gpc \cite{2016MNRAS.457.4340K}. It assumes a Planck cosmology with $\Omega_m = 0.307$, $\Omega_{\Lambda} = 0.693$ $\sigma_8 = 0.823$, and $H_0 = 67.7$ km s$^{-1}$ Mpc$^{-1}$. With SDSS and DESI data in mind, we select a simulation snapshot with a redshift representative of the median redshift of SDSS \texttt{redMaPPer} clusters (\cite{redmapper}, \cite{redmapper2}) and DESI Bright Galaxy Sample galaxies of $z \approx 0.194$, corresponding to a scale factor of $a=0.83760$. Dark matter halos are identified using the \texttt{Rockstar} halo-finder algorithm (\cite{2013ApJ...762..109B}) and merger trees are built from the Consistent-Tree algorithm (\cite{2013ApJ...763...18B}). 

MDPL2 halos are populated with galaxies using the \texttt{UniverseMachine} algorithm (\cite{2019MNRAS.488.3143B}). Briefly, \universemachine\ parametrizes the halo's star formation rate in terms of the halo's growth rate, varying the model parameters so as to match a variety of properties of the galaxy distribution across cosmic times. We impose a stellar mass cut of $M_{*} = 10^{10}  \ h^{-1}M_{\odot} $, and further restrict our analysis to massive halos with mass $M_{\rm vir} \geq 8\times10^{13} \ h^{-1} M_{\odot}$, defining $M_{\rm vir} = \frac{4\pi}{3} r^3_{\rm vir} \Delta \rho_c $ as in \cite{1995AAS...187.9504B}.

Following the methodology of \cite{10.1093/mnras/staa3994},  we employ the distant observer approximation and take the $z$-axis of the simulation box as our line of sight (LOS). The three-dimensional and projected radial separation of galaxies relative to the central galaxy of each halo are given by
\begin{equation}
  r =\sqrt{(x-x_{\rm cen})^2 + (y-y_{\rm cen})^2 + (z-z_{\rm cen})^2} , 
\end{equation}
and
\begin{equation}
R =\sqrt{(x-x_{\rm cen})^2 + (y-y_{\rm cen})^2}
\end{equation}
respectively. Their relative LOS velocities are:
\begin{equation}
    v_{\mathrm{LOS}} = v_{\mathrm{pec, LOS}} \pm aHr_{\mathrm{LOS}},
\end{equation}
where $v_{\mathrm{pec, LOS}}=v_z-v_{z, \rm cen}$, and $r_{\mathrm{LOS}} = \sqrt{r^2-R^2}$ is the co-moving LOS separation. The $\pm$ sign depends on whether the galaxy is in front (-) or behind (+) the galaxy cluster, i.e. the Hubble velocity contribution is always away from the cluster.  When constructing our synthetic galaxy catalog, we limit our selection to galaxies with $R < 30 \ h^{-1}$ Mpc and $|v_{\mathrm{LOS}}|<4,000$ km/s. We bin cluster-galaxy pairs by halo mass in logarithmically-spaced bins with $M_{\rm vir} \in [0.8, 16.0]\times 10^{14}\ h^{-1} M_\odot$, joining the highest two bins due to the low number of halos in the highest bin. Lastly, we bin pairs by cluster centric three-dimensional and projected radius, with radii ranging between $5\ h^{-1}$~Mpc and $30\ h^{-1}$~Mpc in bins of width $1\ h^{-1}$~Mpc. The innermost bin edge is the smallest radius to have no orbiting galaxies for halos of any mass (see Fig. \ref{fig:cf}). 

\section{The Velocity Distribution of Infalling Galaxies} \label{sec:VD}

Our goal is to characterize the LOS velocity distribution of infalling galaxies outside the halo, which is itself a projection of the three dimensional velocity distribution $P(\vec v|M,R)$. Consequently, we begin by characterizing this latter probability distribution.  Specifically, we separate $\vec v$ into its radial $(\vr)$ and tangential $(\vt)$ components.  Here, we follow the conventions of \cite{zuweinberg13}, and refer to $\vt$ as the component of the tangential velocity that is \it not \rm orthogonal to the line of sight.  That is, on average, $\vt$ is really only ``half'' of the tangential velocity.  We refer to Fig.~2 in \citet{zuweinberg13} for a helpful illustration of the system's geometry.

As emphasized in \citet{zuweinberg13}, \textit{the velocity distribution $P(\vr,\vt)$ is not separable}. To simplify, we decompose this distribution in two using conditional probability distributions, i.e.
\begin{equation}
    P(\vr,\vt|r) = P(\vr|r)P(\vt|\vr, r),
\end{equation} \label{eqn:joint}
where we have left the mass dependence of these distributions implicit.  We set out to characterize the radial and tangential velocity distributions above.

\subsection{The Distribution of Radial Velocities  \texorpdfstring{$P\left(v_{\rm r}|r\right)$}{TEXT}  }
\label{sec:rad}

\begin{figure*}
    \centering
    \includegraphics[width=\textwidth, height=8cm]{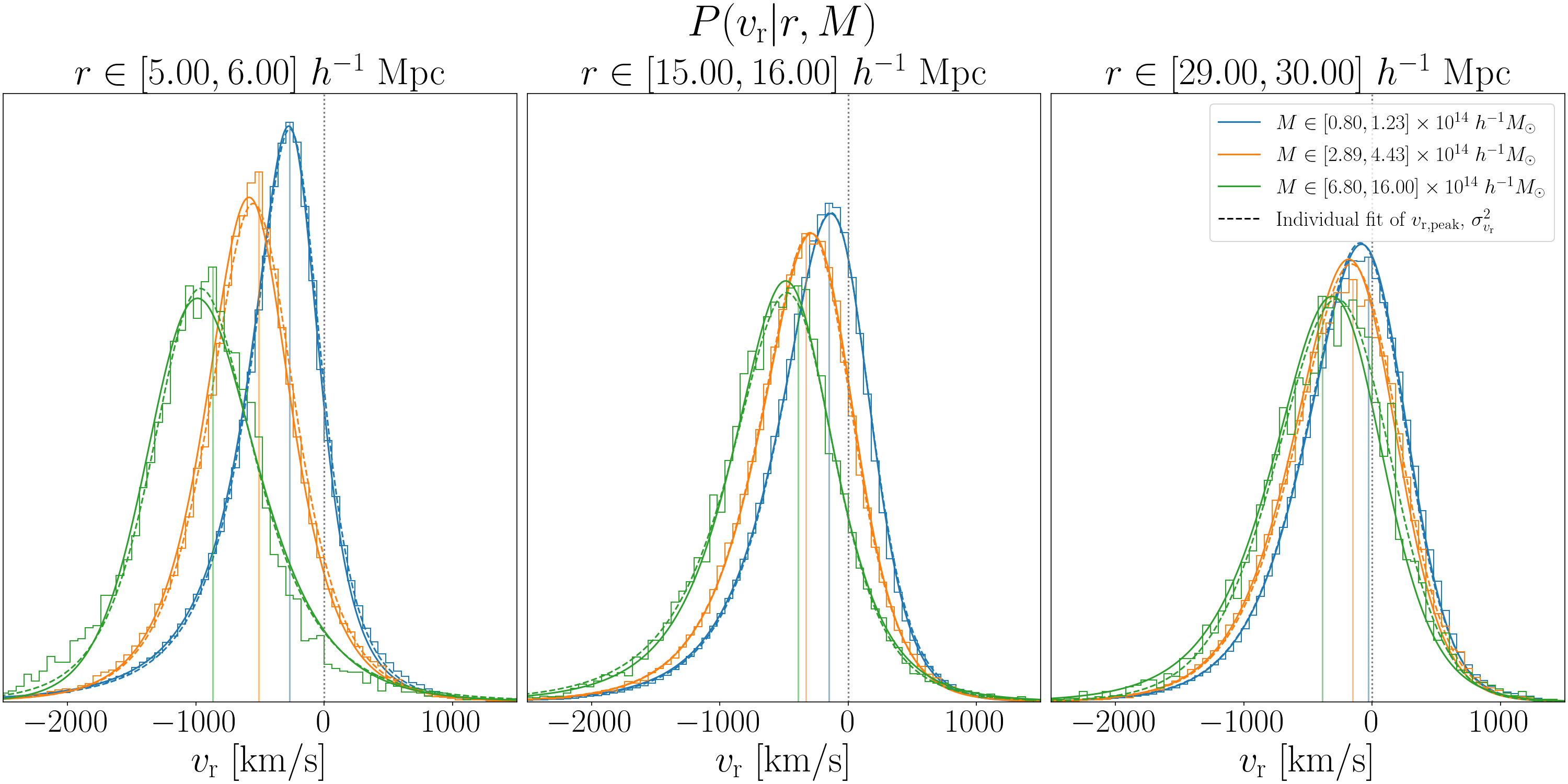}
    \caption{Marginal distributions of radial velocities in several narrow radial bins and several halo mass bins. For each we show fits to the individual bin using the two parameter JSU model with the dashed curves and the best-fit model accounting for mass dependence outlined in ~\ref{sec:rad} in the solid curves. The vertical lines indicate the peaks of each distribution and $\vr = 0$.}  
    \label{fig:rad_marg_dists_indiv}
\end{figure*}

Fig.~\ref{fig:rad_marg_dists_indiv} shows the marginal radial velocity distributions $P(v_{\rm r}|r)$ for several radial and halo mass bins. The distributions peak at negative $v_{\rm r}$ for all scales probed, and exhibit both skewness and large tails (kurtosis). We have found that the Johnson's SU (JSU) distribution used in \cite{10.1093/mnras/stz2227} can accurately fit the distributions in the simulation. This distribution $\rho_{\rm JSU}(x)$ is defined by assuming the transformed variable $z=\gamma+\delta\sinh^{-1}\left(\frac{x-\xi_{\rm JSU}}{\lambda}\right)$ is normally distributed, i.e. $z\sim\mathcal{N}(0,1)$. For our purposes, $x=\vr$, and $\gamma, \ \delta, \ \xi_{\rm JSU}$ and $\lambda$ are the four parameters governing the mean, variance, skewness, and kurtosis of the distribution, though these parameters don't map one-to-one onto these statistics. 

We fit the distribution of radial velocities for each of our mass and radial bins using the JSU distribution.  Remarkably, we find that while the JSU parameters vary with both radius and mass, the best fit values cluster in lines in parameter space, enabling us to write the ``dimensionless'' parameters JSU parameters $\gamma$ and $\delta$ as linear functions of the ``dimensionfull'' parameters $\xijsu$ and $\lambda$.  That is, we set
\begin{align}
    \bar \gamma(\xi_{\rm JSU}) &= (7.8\times10^{-4}\ \text{km}^{-1}\text{s})\xi_{\rm JSU}+0.5859\label{eqn:bargam} \\
    \bar \delta(\lambda) &= (2.09\times10^{-3}\ \text{km}^{-1}\text{s})\lambda + 0.2870  \label{eqn:bardelt}
\end{align}
These relations govern the shape of the velocity probability distribution, enabling us to reduce the number of degrees of freedom in our fits from four to two, namely $\xijsu$ and $\lambda$.

In practice, we find it more useful to recast these variables in terms of more interpretable parameters. Specifically, we define $\vrpk$ as the most probable infall radial velocity, i.e. the velocity corresponding to the peak of the distribution $P(\vr|r)$, and $\svr$, the variance of this distribution.  To do so, we simply evaluate $\vrpk$ and $\svr$ for a broad range of $\xijsu$ and $\lambda$ values, and then build an interpolator\footnote{We use Scipy's RBFInterpolator for this task.} that enables us to quickly calculate $\xijsu$ and $\lambda$ as a function of $\vrpk$ and $\svr$. In this way, we are able to fit the velocity probability distributions in terms of these last two interpretable parameters. We do so using the Poisson likelihood:
\begin{equation}\label{eqn:Poisson}
    \ln\mathcal{L} = \sum_{r, v_{\rm r}} n_{ij} \ln(\lambda_{ij}) -\lambda_{ij} - \ln(n_{ij}!),
\end{equation}
where  $n_{ij}$ is the number of galaxies in a radial velocity bin for a given radial bin, $\lambda_{ij}=P(v_{r, i}|r)\delta_{v_{\rm r}}N$, $\delta_{v_{\rm r}}$ is the width of the radial velocity bin, and $N$ is the number of galaxies in the radial bin. This follows the assumption of \cite{2024MNRAS.533.4081R} that the number of galaxies in a velocity bin are Poisson distributed. For our ML estimation, we use 50 velocity bins in the range $v \in [-3,000, 3,000]$~km/s.  We obtain parameter posteriors with the MCMC library EMCEE \cite{emcee} using flat priors on all of the parameters, only enforcing that parameters keep their initial sign.  

Figure~\ref{fig:rad_marg_dists_indiv} compares the velocity distributions $P(\vr|r)$ in the simulations (colored histograms) to our best fit models using only the two degrees of freedom $\vrpk$ and $\svr$ (dashed curves).  We can see that our model provides an excellent fit to the data. 

We now parameterize the radius and mass dependence of the velocity distribution.  Starting with the radial dependence, we find that the peaks and variances are well described by the power laws: 
\begin{align}
    v_{r, \rm peak}(r) &= v_{r, p}\left(\frac{r}{r_p}\right)^{v_{r, s}} \\
    \sigma_{v_{\rm r}}^2(r) &= \sigma^2_{r, p}\left(\frac{r}{r_p}\right)^{\sigma_{r, s}}+\sigma^2_{r, c}
\end{align}
where we set a pivot scale of $r_p=10 \ h^{-1}$ Mpc. 

\begin{figure*}
    \centering
    \includegraphics[width=\textwidth, height=9cm]{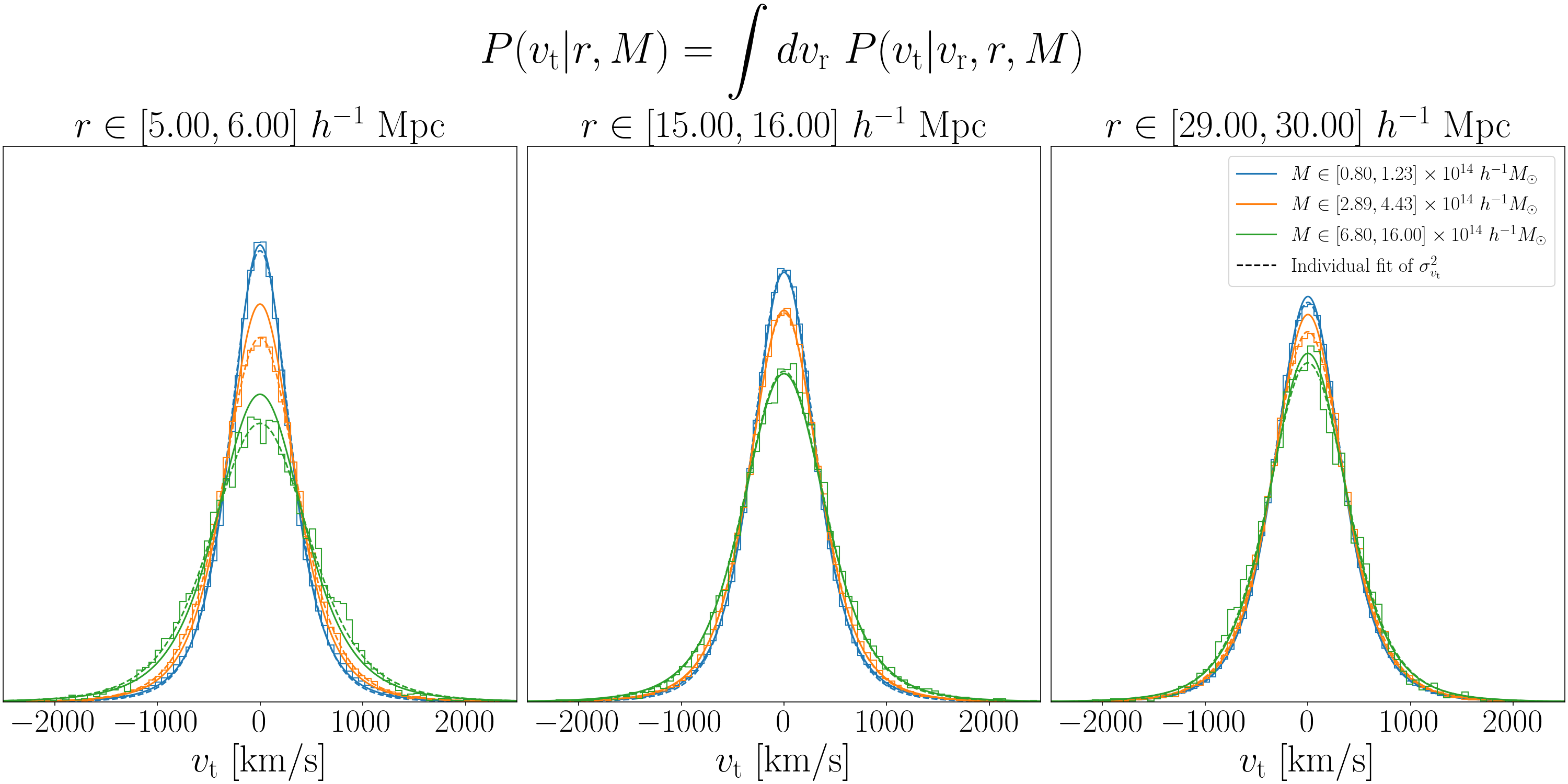}
    \caption{Distributions of the marginal tangential velocities in distributions in several narrow radial bins and halo mass bins. For each we show with a smooth curve the result from marginalizing over the best-fit model outlined in Section ~\ref{sec:tan}, and with a dashed curve that from directly fitting the variance of the conditional distributions.} 
    \label{fig:tan_cond_dists_indiv}
\end{figure*}

We wish to further characterize the mass dependence of the probability distribution $P(\vr|r,M)$.  To that end, we inspect how the parameters of the above power laws vary as functions of mass. In fitting the mass dependence of our model parameters, we assigned to each mass bin a single mass equal to the median mass of the halos within the bin. We find that all but $\sigma_{r, c}$ follow power laws of the form 
\begin{align}
    \theta_x(M) = \theta_{x, p}\left ( \frac{M}{M_p} \right)^{\theta_{x, s}},
\end{align}
where we use a pivot mass of $M_p = 10^{14} \ \mathrm{M}_\odot$. The large-scale limit of the radial galaxy velocity dispersion $\sigma^2_{r,c}$ is mass independent (far from the halos, the velocity dispersion doesn't depend on halo mass), for which we find $\sigma^2_{r, c}=2.13\times10^{5}\ (\text{km/s})^2$. We refer to the resulting fit for $P(\vr|r,M)$ as our \textit{smooth model fit}.

When comparing the quality of our bin-by-bin fits (dashed lines in Fig.~\ref{fig:rad_marg_dists_indiv}) to those of our smooth model fit, we find that the shape of our smooth-model distribution differs sufficiently from the data to warrant modifying our model.  Comparing our smooth model parameters to the bin-by-bin best fit models, we find that the largest differences are those of the intercept of $\bar \delta(\lambda)$, particularly at high mass. Consequently, we introduce an additional mass dependence for the model parameter $\delta$ of the form
\begin{equation}
    \delta(\lambda)=\bar \delta(\lambda) + \Delta\delta(M),
\end{equation}
where
\begin{equation}
    \Delta\delta(M) = \Delta_m \left(\frac{M}{M_p}\right) + \Delta_b.
\end{equation}
The quantities $\Delta_m$ and $\Delta_b$ are model parameters that we can use to further improve the quality of our smooth model fit.  The resulting best fit smooth model for $P(v_{\rm r}|r, M)$ is shown as the solid curves in Fig.~\ref{fig:rad_marg_dists_indiv}.  We can see our final smooth model broadly reproduces our bin-by-bin fits, capturing both the peak position and skewed tails of the distribution, features that are essential for constraining cluster masses in the manner of \cite{10.1093/mnras/stz2227}.

\begin{figure*}
    \centering
    \includegraphics[width=\textwidth, height=16cm]{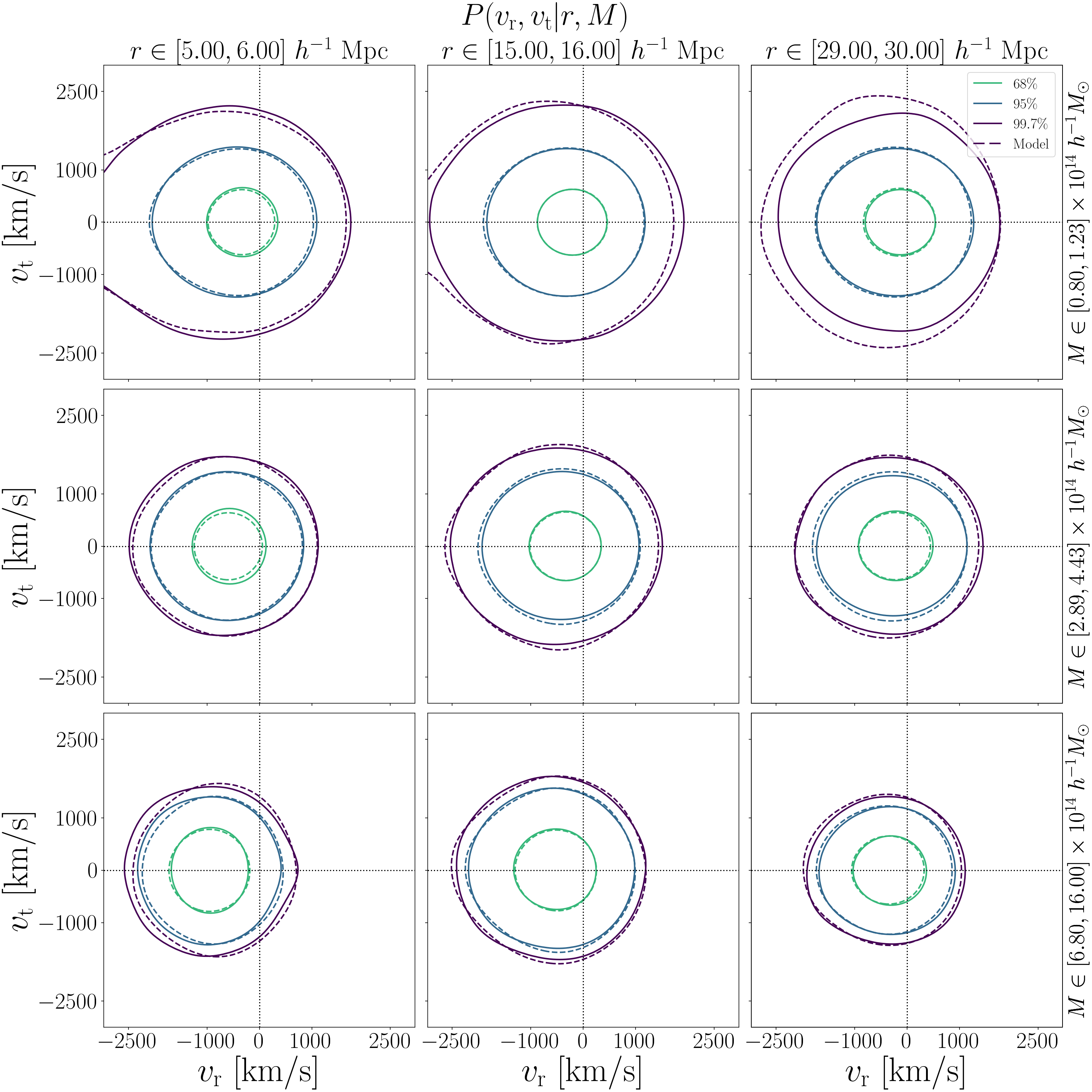}
    \caption{Comparison of the $68\%$, $95\%$, and $97\%$ intervals of the joint distributions (solid contours) and the smooth model (dashed contours) outlined in Section~\ref{sec:VD}, smoothed with a gaussian filter. We show the joint distributions for three radial bins (columns) and three halo mass bins (rows). Black dotted lines indicate $v_{\rm r}, v_{\rm t}=0$. }
    \label{fig:joint_cond_dists_smooth}
\end{figure*}

\subsection{The Conditional Distribution of Tangential Velocities  \texorpdfstring{$P\left(v_{\rm t} |v_{\rm r},r\right)$}{TEXT}} \label{sec:tan}

We now turn to modeling the conditional probability distributions $P(\vt|\vr,r)$. These distributions are necessarily symmetric around $\vt=0$, and exhibit large tails. Following \cite{zuweinberg13}, we model these distributions by assuming the quantity $\vt/\sigma_{\vt}$ is distributed as a Student's t-distribution for a judiciously chosen value of $\sigma_{\vt}$. We find that we can set the degrees-of-freedom of the t-distribution to $\texttt{dof}=5$ independent of mass and radius, consistent with the findings of \cite{PhysRevD.70.083007} (see their Fig. 3). By contrast, the parameter $\sigma_{v_{\rm t}}$ does vary with both $\vr$, $r$, and halo mass. 

We perform bin-by-bin fits of $P(\vt)$ in bins of radius, radial velocity $\vr$, and mass.  At fixed mass and radius,  $\sigma_{v_{\rm t}}$ exhibits a clear asymmetric minimum as a function of $\vr$, leading us to model this dependence as a cubic of the form\footnote{As a cubic, our fitting function does not extrapolate well, becoming negative at large negative radial velocities. When performing numerical integrals, we linearly extrapolate beyond the minimum and maximum radial velocities used to calibrate $\sigma_{\vt}^2$, enforcing continuity of the function and its derivative at the endpoints.}
\begin{equation}\label{eqn:cubic}
    \sigma_{v_{\rm t}}^2(v_{\rm r}|r) = A (v_{\rm r}-\mu)^3 + B(v_{\rm r}-\mu)^2 + C, 
\end{equation}
where $A$, $B$, $C$, and $\mu$ are model parameters.  Upon fitting the simulation data, we find that $A$ and $B$ do not depend on radius, but $C$ and $\mu$ do.  These dependencies can be fit using the fitting functions
\begin{align}
    \mu(r) &= \mu_0 + \mu_1/r \label{eqn:sig_m} \\
    C(r) &= C_0 + C_1(r-r_C) \label{eqn:sig_0}.
\end{align}
In these expressions, $\mu_0$, $\mu_1$, $C_0$, $C_1$, and $r_C$ as model parameters which may depend on mass.\footnote{Note our model is only valid over the radial range in which it has been calibrated: native interpolation to large radii can result in negative $C$ values, which is unphysical.}

We now have a smooth parameterization for $P(\vt|\vr,r)$ in terms of the model parameters $A$, $B$, $\mu_0$, $\mu_1$, $C_0$, $C_1$, and $r_C$.  We fit the distribution for each of the mass bins in our simulation data to determine the mass evolution of our model parameters. We find $A$ does not depend on mass, while $B$ varies as a simple power-law
\begin{equation}
    B(M)=B_p\left(\frac{M}{M_p}\right)^{B_s}.
\end{equation}
The expressions for $\mu_0$, $\mu_1$, and $C_1$ are different, since $\mu_0$ is generally negative and $\mu_1$ and $C_1$ change sign as a function of mass.  Consequently, we model $\mu_0(M)$, $\mu_1(M)$, and $C_1(M)$ as linear functions in $M$,
\begin{align}
    \mu_0(M) &= \mu_{0, c} + \mu_{0, p}(M/M_p) \\
    \mu_1(M) &= \mu_{1, c} + \mu_{1, p}(M/M_p) \\
    C_1(M) &= C_{1, c}+C_{1, p}(M/M_p),
\end{align}
where the parameters $\theta_{x, c}$, $\theta_{x, p}$ are mass independent.

To illustrate the quality of our fits, in Fig.~\ref{fig:tan_cond_dists_indiv} we compare the observed marginal distributions
\begin{equation}
    P(\vt|r,M) = \int d\vr\ P(\vt|\vr,r,M)P(\vr|r,M)
\end{equation}
from the simulation data to that of our best fit model.  We can see that our model provides an accurate description of the marginal probability distribution functions. 

\begin{table*}
    \centering
    \caption{Maximum a posteriori values of the parameters for the $P(v_{\rm r}, v_{\rm t}|r, M)$ model along with $1\sigma$ intervals.}
    \label{tab:joint_params}
    \begin{tabular}{|c|c|c|}
        \hline
        Parameter & Description & MAP$\pm1\sigma$ \\ 
        \hline
        $v_{p, p}$ & Infall velocity power law pivot [km/s] & $188.71 \pm 0.16$ \\ 
        $v_{p, s}$ & --- & $0.57518 \pm 0.00084$ \\ 
        $v_{s, p}$ & Infall velocity power law slope & $0.6301 \pm 0.0013$ \\ 
        $v_{s, s}$ & --- & $0.0685 \pm 0.0020$ \\ 
        \hline
        $\sigma_{p, p}$ & Radial velocity variance power law pivots [(km/s)$^2$] & $193174 \pm 15$ \\ 
        $\sigma_{p, s}$ & --- & $-0.87735 \pm 0.00029$ \\ 
        $\sigma_{s, p}$ & Radial velocity variance slope & $1.17179 \pm 0.00015$ \\ 
        $\sigma_{s, s}$ & --- & $-0.27437 \pm 0.00038$ \\ 
        \hline
        $\Delta_m$ & Radial velocity shape mass dependence linear variation & $-0.03961 \pm 0.00024$ \\ 
        $\Delta_b$ & --- & $0.39943 \pm 0.00046$ \\ 
        \hline
        $A$ & Cubic term amplitude of $v_{\rm t}|v_{\rm r}$ variance [(km/s)$^{-1}$] & $0.00004206 \pm 0.0000043$ \\ 
        $B_p$ & Quadratic term amplitude of $v_{\rm t}|v_{\rm r}$ variance & $0.17355 \pm 0.00044$ \\ 
        $B_s$ & --- & $-0.1024 \pm 0.0030$ \\ 
        $\mu_{0,p}$ & Minimum of $v_{\rm t}|v_{\rm r}$ variance [km/s] & $-22.58 \pm 0.76$ \\ 
        $\mu_{0,c}$ & [km/s] & $63.2 \pm 1.7$ \\ 
        $\mu_{1,p}$ & [$h^{-1}$ Mpc km/s ] & $-508.9 \pm 11.3$ \\ 
        $\mu_{1,c}$ & [$h^{-1}$ Mpc km/s] & $-1094 \pm 22$ \\ 
        $C_{1,p}$ & Large-scale constant of $v_{\rm t}|v_{\rm r}$ variance [$h$ Mpc$^{-1}$ km/s] & $-268.4 \pm 1.7$ \\ 
        $C_{1,c}$ & [$h$ Mpc$^{-1}$ km/s] & $1108.0 \pm 2.9$ \\ 
        \hline
    \end{tabular}
\end{table*}

\subsection{The Joint Distribution of Peculiar Velocities} \label{sec:JD}

Fig. ~\ref{fig:joint_cond_dists_smooth} shows the result of computing ~\ref{eqn:joint}
with our best fit smooth model on a grid of radial and mass bins. We compare the $68\%$, $95\%$, and $99.7\%$ probability regions for the model and data in each bin, smoothed by a gaussian filter, finding good agreement between the two generally. We list the MAP values for each of our smooth model parameters in Table \ref{tab:joint_params}. We note that the greatest deviations between the model and the data appear for the combination of our lowest radial bin and highest mass bin. This mass bin in particular is the most susceptible to contamination from orbiting galaxies at our innermost radial bin.  This source of bias can be easily removed by moving the boundary to slightly larger radii for the most massive halos.  Since this is the least populated bin, the relative decrease in signal-to-noise relative to the whole sample is very low.  For now, we simply adopt this 5~$h^{-1}$ Mpc cut as our fiducial value. Overall, we find good agreement within the 95\% probability region.


\section{The Halo-Galaxy Correlation Function}\label{sec:cf}
\begin{figure}
    \includegraphics[width=\columnwidth, height=6cm]{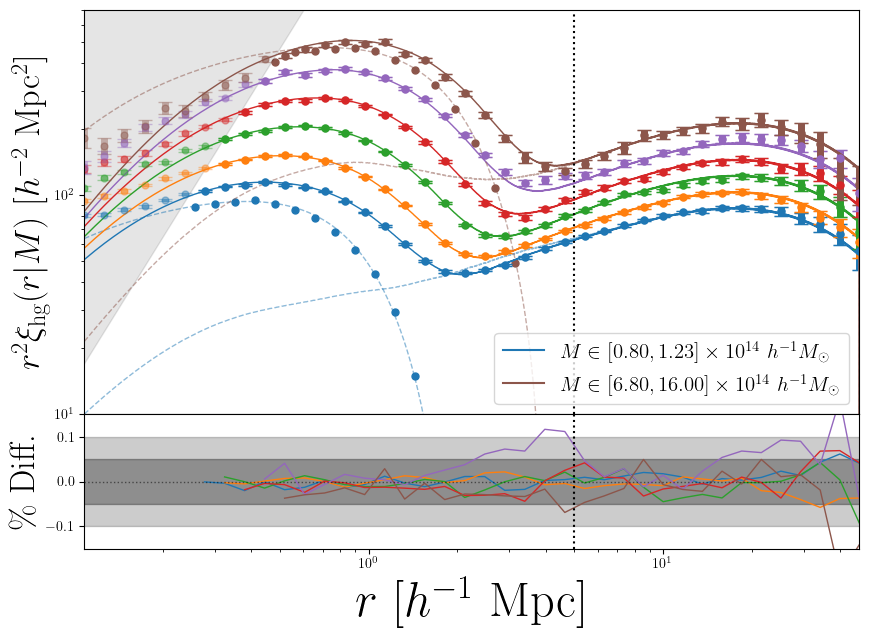}
    \caption{ \textbf{Top:} The halo-galaxy correlation function measured for MDPL2 Rockstar halos and UniverseMachine galaxies in each of our halo mass bins. Uncertainties are computed by jackknifing halos within sub-boxes of the simulation box. The solid curves show our best-fit of the \cite{PhysRevD.111.043527} model for total halo-galaxy correlation function. The orbiting and infalling components of this model for the lowest and highest mass bins are shown with dashed curves. Additionally, the corresponding measured orbiting components are shown as points without uncertainties. The minimum radial bin edge used for our analysis of the velocity distributions is indicated by the vertical black dotted line.  \textbf{Bottom:} Percent difference between data and model for $\xi_{\rm hg}(r|M)$. The dark and light gray bands indicate differences with $5\%$ and $10\%$, respectively. }
    \label{fig:cf}
\end{figure}

The distribution of LOS velocities at a given radius is determined in part by a weighted average of the joint distribution of velocities $P(v_{\rm r}, v_{\rm t}|r)$ over LOS distance, where the relative weighting function is the density profile of galaxies $\rho(r)$. In our case, this is just the density profile of \it infalling \rm galaxies. To avoid contamination from orbiting galaxies, we set the smallest scale used for our study of the velocity distributions so that the density profile receives negligible contributions from orbiting galaxies. We highlight this scale in Fig. ~\ref{fig:cf}, in comparison to the orbiting and infalling components of the total density profile.  

We adopt the same distinction between `orbiting' and `infalling' galaxies found in \cite{PhysRevD.111.043527}, and also utilize their model for the infalling density profile. This component of the model is described in terms of the infalling correlation, which is expressed as a biased form of the correlation function at large scales:
\begin{equation}
    \xi_{\mathrm{inf}}(r|M) = \frac{\beta(r)}{1+\eta x\exp(-x)}b\xi_{\mathrm{LS}},
\end{equation}

where $x\equiv r/r_{\mathrm{h}}$. This $r_{\mathrm{h}}$ is the physical scale associated with the exponential cut-off and the extent of orbiting galaxies. The large scale correlation function is chosen to be $\xi_{\rm{LS}} \equiv B\xi_{\mathrm{ZA}}$, where $B=0.948$ is a simulation-calibrated constant, and $\xi_{\mathrm{ZA}}$ is the correlation function in the Zel'dovich approximation. The scale-dependent bias $\beta(r)$ impacts small scales, and is written:
\begin{equation}
    \beta(r)=1+\left[\frac{r_{\rm inf}}{\mu r_{\mathrm{h}}+r}\right]^\gamma.
\end{equation}
$r_{\rm inf}$ is a characteristic nonlinear scale and, along with $\mu$, does not vary with mass. The edge radius $r_{\rm h}$,  scale-dependent bias slope $\gamma$, infalling profile `dip' amplitude $\eta$, and the large scale bias parameter $b$ vary with mass:
\begin{align}
    r_{\rm h} &= r_{\rm{h}, p} \left ( \frac{M}{M_p} \right )^{r_{\rm{h}, s}} \\
    \eta &= \frac{1}{2}\eta_0\left[1-\mathrm{erf} \left\{\frac{\log(M/M_p)-\eta_m}{\eta_\sigma}\right]\right\}\\
    \gamma &= \gamma_p\left(\frac{M}{M_p}\right)^{\gamma_s}\\
    b &= b_p\left(\frac{M}{M_p}\right)^{b_s}.
\end{align}

With each of these models, we fit the respective component's profile within each of our specified mass bins simultaneously. The resulting infalling halo-galaxy correlation function is shown in Fig. ~\ref{fig:cf}. We also show the total halo-galaxy correlation function defined in \cite{PhysRevD.111.043527} computed as $\xi_{\rm hg}(r|M)=\left(\frac{\rho_{\rm orb}(r|M)}{\bar \rho_g}-1\right) + \xi_{\rm inf}(r|M)$. We measure the average density of galaxies in the simulation to be $\bar \rho_{g} = 0.014$ $h^3$Mpc$^{-3}$. 

For the density profiles, we assume the likelihood

\begin{equation}
    \ln \mathcal{L} = -\frac{1}{2}\sum_{r, M} \left[\frac{(\rho^{\mathrm{data}}_{ij}-\rho^{\mathrm{model}}_{ij})^2}{s^2_{ij}}+\ln(s^2_{ij}) \right],  
\end{equation}
where
\begin{equation}
    s^2_{ij}=\sigma^2_{ij}+\Delta^2(\rho^{\mathrm{model}}_{ij})^2.
\end{equation}
$\sigma^2_{ij}$ is the data uncertainty, which we compute by jackknifing halos in sub-boxes. The latter term corresponds to an inherent scatter $\Delta$ in the data about the expected relation determined by the model. We obtain parameter posteriors as in ~\ref{sec:rad}; The maximum a posteriori values are reported in Table \ref{tab:hg_params_inf}. We find a best fit model scatter $\Delta = 2.12\%$. Based on our best fit model, we can see that our fiducial radial cut of $r\geq 5\ h^{-1}\ {\rm Mpc}$ can excise the contribution of orbiting galaxies to the velocity dispersion of galaxies along the line of sight.

\begin{table}
    \centering
    \caption{Maximum a posteriori values of the parameters of the infalling halo-galaxy correlation model. We include the $1\sigma$ uncertainties, reporting that bound with the largest magnitude.}
    \label{tab:hg_params_inf}
    \begin{tabular}{|c|c|}
        \hline
        Parameter & $\xi_{\rm inf}$ \\ 
        \hline
        $r_{\mathrm{h}, p}$ & $0.735\pm0.014$ [$h^{-1}$ Mpc]\\
        $r_{\mathrm{h}, s}$ & $0.281\pm0.012$ \\ 
        $b_p$ & $5.380 \pm 0.035$ \\ 
        $b_s$ & $0.4225 \pm 0.0058$ \\ 
        $\gamma_p$ & $2.523 \pm 0.085$ \\ 
        $\gamma_s$ & $0.064 \pm 0.013$ \\ 
        $\eta_0$ & $2.80 \pm 0.47$ \\ 
        $\eta_m$ & $0.447 \pm 0.026$ \\ 
        $\eta_\sigma$ & $0.381 \pm 0.034$ \\ 
        $r_{\rm inf}$ & $2.98 \pm 0.13$ [$h^{-1}$ Mpc] \\ 
        $\mu$ & $1.005 \pm 0.085$ \\
        \hline
        $\Delta$ & $0.021213 \pm 0.010000$ \\ 
        \hline
    \end{tabular}
\end{table}


\section{The Distribution of Line-of-Sight Velocities}\label{sec:LOSVD}

\begin{figure*}
    \centering
    \includegraphics[width=\textwidth, height=10cm]{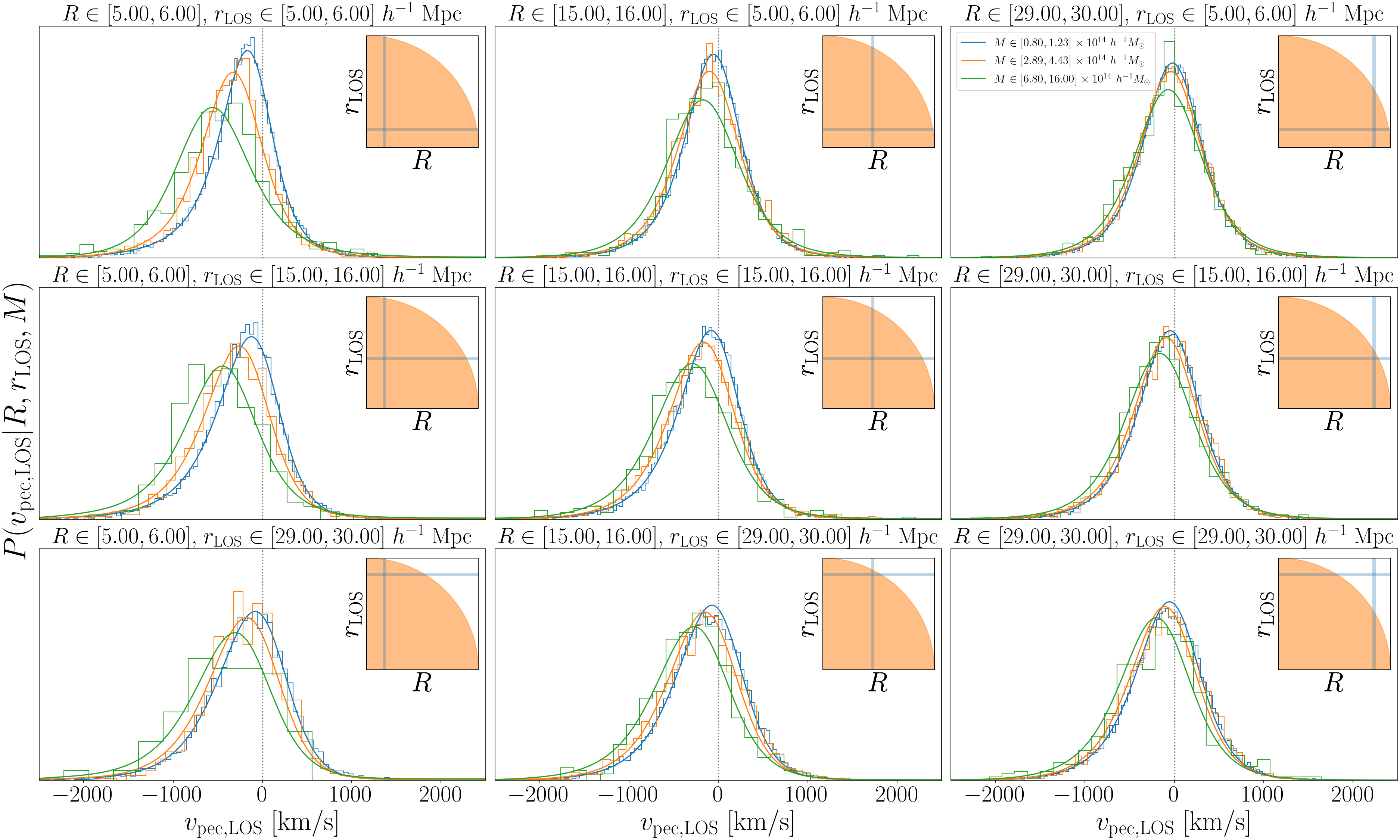}
    \caption{The distribution of peculiar LOS velocities in several bins of projected radius and LOS separation in our lowest, central and highest mass bins, inspired by Fig. 7 of \cite{zuweinberg13}. Curves show the result of the integration in equation ~\ref{eqn:Pvlos_2d} and vertical dotted lines indicate $v_{\rm r}=0$. 
    }
    \label{fig:Pvlos}
\end{figure*}

With our models for $P(v_{\rm r}, v_{\rm t}|r, M)$ and $\rho_{\rm inf}(r|M)$ fully specified, we proceed to computing the LOS velocity distributions. The peculiar LOS velocity distribution at a given projected radius and LOS distance $P(\vpeclos | R, \rlos)$ is the integral of the joint distribution over each component, which we can write 
\begin{equation}\label{eqn:Pvlos_2d}
    P(\vpeclos | R, \rlos) = \int \frac{dv_{\rm r}}{\cos\theta} P(\vr, \vt = \vt' | r ), 
\end{equation}
where we maintain the conventions of \cite{zuweinberg13}, with $\vt' = \frac{\vpeclos - \vr \sin\theta}{\cos\theta}$ and $\theta = \tan^{-1}(r_{\mathrm{LOS}}/R)$. The distribution of physical LOS velocities is simply obtained by taking $\vpeclos=\vlos - aH\rlos$. Integrating along the LOS, we have that the projected LOS velocity distribution is 
\begin{equation}\label{eqn:Pvlos_1d}
    P(\vlos|R) = \frac{ \displaystyle \int dr_{\mathrm{LOS}}\, P(\vlos|R, \rlos)\rho_{\rm inf}(r)}{\Sigma_{\rm inf}(R)},
\end{equation}
where $\Sigma_{\rm inf}(R)$ is the projected surface density profile. This profile is the result of integrating the density profile along the LOS while accounting for the noise due to peculiar velocities:
\begin{align}
    \Sigma_{\rm inf}(R) &= \int dr_{\mathrm{LOS}}\, \rho_{\rm inf}(r) P(|v_{\mathrm{LOS}}| \leq v_{\rm max}) \\                  
    & = \frac{1}{2}\int dr_{\mathrm{LOS}}\, \rho_{\rm inf}(r)\left[1+\mathrm{erf}\left(\frac{v_{\rm max} -Hr_{\mathrm{LOS}}}{\sqrt{2}\sigma_{\mathrm{LOS}}}\right)\right], 
\end{align}
where we have computed $P(|v_{\mathrm{LOS}}|\leq v_{\rm max})$ assuming that $P(v_{\mathrm{LOS}})$ appears Gaussian far from the cluster.  The LOS velocity dispersion value is measured directly from the simulation to be $\sigma_{\mathrm{LOS}}^2=(532\ {\rm km/s})^2$. 

Putting these expressions together, we have that the projected LOS velocity distributions are 
\begin{equation}
    P(\vlos|R) = \frac{\displaystyle \int d\rlos \int \frac{d\vr}{\cos\theta} P(\vr, \vt=\vt'|r)\rho_{\rm inf}(r) }{\Sigma_{\rm inf}(R)} .
\end{equation}
All of the integrals discussed in this section make use of finite integration limits $|v_i|<2,500$ km/s and $|r_{\mathrm{LOS}}|<40 \ h^{-1}$Mpc, which we have found is sufficient for convergence over the range of velocities and projected radii employed in this work. We also utilize integrands evaluated on precomputed grids of $(v_{\rm r}, v_{\rm t})$, $r_{\mathrm{LOS}}$ for expediency, without compromising significant numerical accuracy\footnote{Our integration of the joint distribution, implemented with SciPy, is motivated by the approach in \cite{2024MNRAS.533.4081R}.}. 

Fig. ~\ref{fig:Pvlos} shows the results of the integration in equation ~\ref{eqn:Pvlos_2d} for peculiar velocities with our best fit model on a grid of $(R, r_{\mathrm{LOS}})$ for three of our mass bins. We are able to capture the higher order moments and mass dependence of $P(v_{\mathrm{pec, LOS}}|R, r_{\mathrm{LOS}}, M)$, while avoiding the complex dependencies on $R$ and $r_{\mathrm{LOS}}$ that arise when attempting to fit these distributions directly.   

\begin{figure*}
    \centering
    \includegraphics[width=\textwidth, height=8cm]{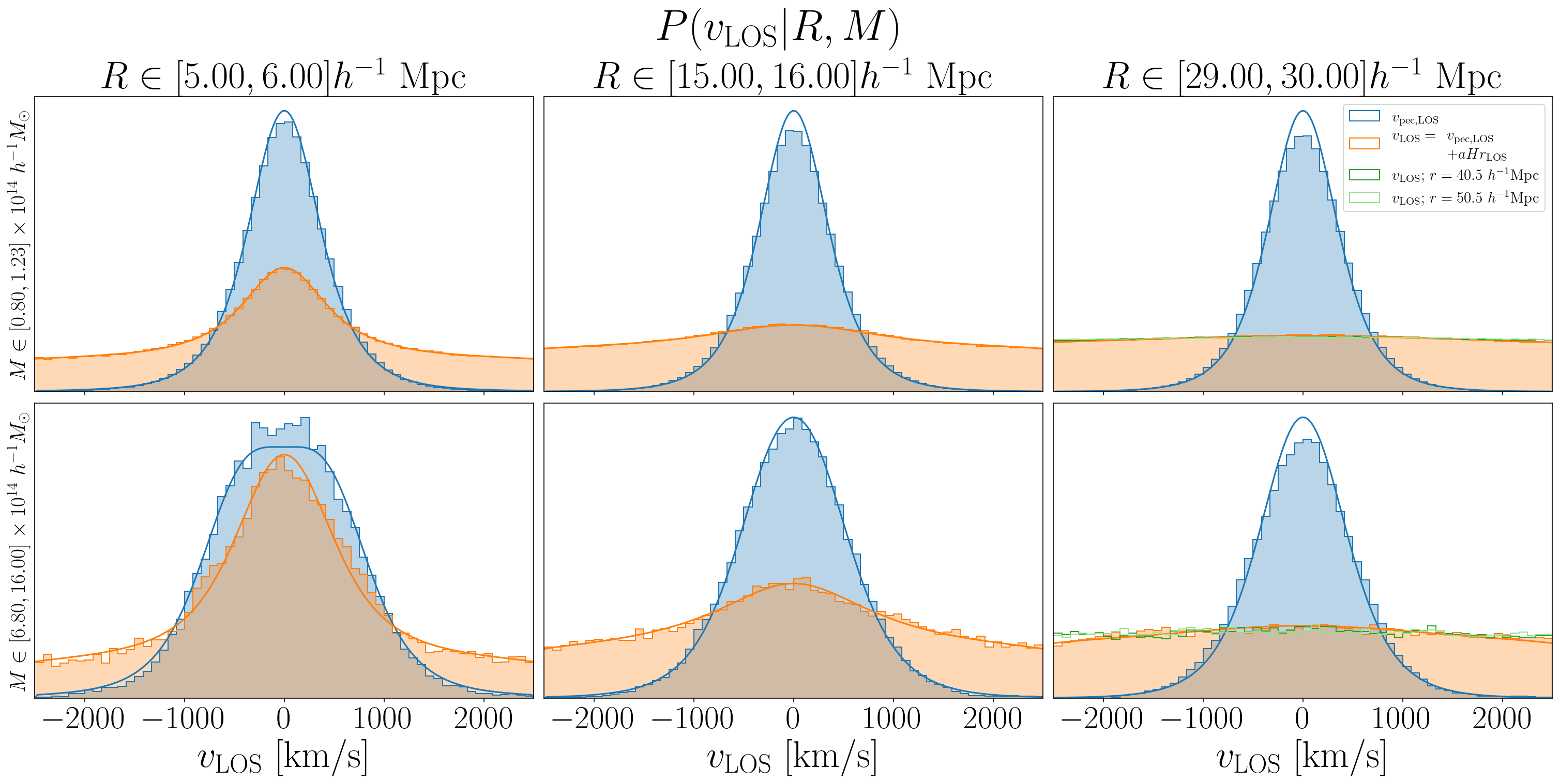}
    \caption{The distribution of peculiar (blue) and physical (orange) LOS velocities in several bins of projected radius $\in [4, 30]h^{-1}$ Mpc in our lowest and highest mass bins. Solid curves show the result of the integration in equation ~\ref{eqn:Pvlos_1d}.  
    }
    \label{fig:Pvphys}
\end{figure*}

The peculiar (blue) and physical (orange) LOS velocity distributions that result from equation ~\ref{eqn:Pvlos_1d} are shown in Fig. ~\ref{fig:Pvphys}. We also show for the highest radial bins histograms of the distributions of $P(v_{\mathrm{LOS}}|R)$ in $1 \ h^{-1}$Mpc thick bins centered on $r=40$ and $r=50$. It is apparent by $r\sim30$ that the features near $v_{\mathrm{LOS}}=0$ due to $P(v_{\mathrm{pec, LOS}}|R)$ are almost entirely washed out by the broadening associated with the Hubble flow. Crucially, our approach is able to predict the projected LOS velocity distributions to within $\sim5\%$ accuracy.   

\section{Calibrating Cluster Masses}\label{sec:calib}
\begin{figure}
    \centering
    \includegraphics[width=\columnwidth]{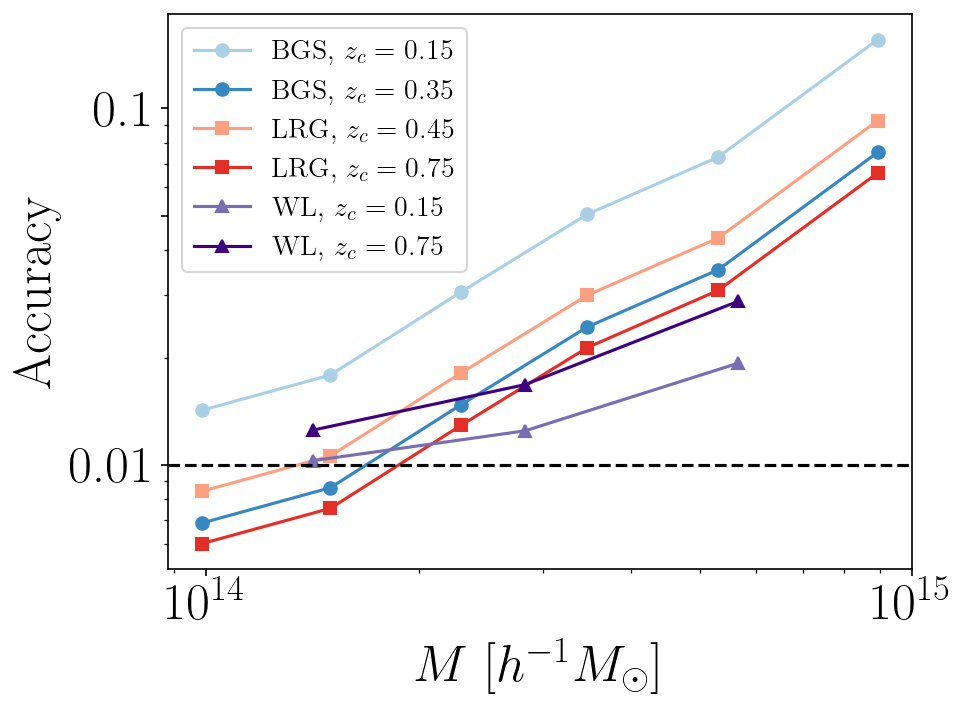}
    \caption{Estimated Fisher mass uncertainties recovered from $P(v_{\rm LOS}|R, M)$ as a function of cluster mass. We compare estimates for DESI BGS and LRG to those for weak lensing stated in \cite{WEINBERG201387}, in narrow redshift bins of $z=z_c\pm0.05$. }
    \label{fig:mass_unc}
\end{figure}

Our model for $P(v_{\mathrm{LOS}} |R, M)$ enables a prediction of the observed number of galaxies in a LOS velocity bin. In a similar manner to \cite{10.1093/mnras/stz2227}, we aim to quantify the sensitivity of this prediction to cluster mass. We forecast the uncertainty of the recovered cluster mass of each of the mass bins used for calibration, while fixing the rest of our model parameters at the cosmology of the simulation snapshot. The prediction for the average number of galaxies in a LOS velocity bin $i$ is 
\begin{equation}
    \langle N (v_{\mathrm{LOS}, i} |R, M) \rangle = N(R, M)\times P(v_{\mathrm{LOS}, i}|R, M)\times \delta v_{\mathrm{LOS}}, 
\end{equation}
where $N(R, M)$ is the total number of galaxies in a bin of $R$ and $M$ and $\delta v_{\mathrm{LOS}}$ is the width of the velocity bins. We then compute the Fisher matrix characterizing the uncertainty in the recovered cluster masses via 
\begin{equation}
    F_{\alpha \beta} =  \sum_{ij} \frac{\partial \langle N (v_{\mathrm{LOS}, i} |R, M) \rangle}{\partial M_\alpha} C_{ij}^{-1}\frac{\partial \langle N (v_{\mathrm{LOS}, j} |R, M) \rangle}{\partial M_\beta}
\end{equation}
with Poisson covariance $C_{ij}=\delta_{ij}N_{ij}$. To estimate the uncertainty on masses constrained with spectroscopic galaxies from DESI, we compute the number of cluster-galaxy pairs as 
\begin{equation}
    N_{\rm pairs, DESI} \sim n_hV_{\rm DESI}\times n_g V_{R},
\end{equation}
where $n_h$ is the number density of halos across the volume of DESI $V_{\rm DESI}$, and $n_g$ is the number density of galaxies in a shell of projected radius around $R$. We can express this in terms of quantities for MDPL2 as
\begin{align}
    N_{\rm pairs, DESI} &= n_h\frac{V_{\rm DESI}}{V_{\rm MDPL2}}V_{\rm MDPL2}\frac{n_g}{n_{g, \rm MDPL2}}n_{g, \rm MDPL2} V_R \\
    &= N_{\rm pairs, MDPL2}\frac{V_{\rm DESI}}{V_{\rm MDPL2}}\frac{n_g}{n_{g, \rm MDPL2}},
\end{align}
where $N_{\rm pairs, MDPL2}$ is the number of pairs in the simulation, and the number density of galaxies in the simulation is the same as $n_{g, \rm MDPL2} = \bar\rho_g=0.014$ $h^3$ Mpc$^{-3}$. We estimate $n_g(z)$ for BGS and LRG galaxies from Fig. 4 of \cite{tr6y-kpc6}, taking $n_{g, \rm BGS}=10^{-3}$ Mpc$^{-3}$ and $n_{g, \rm LRG}=4.5\times10^{-4}$ Mpc$^{-3}$ across the ranges $z\in[0.1, 0.4)$, $z\in[0.4, 0.8)$, respectively. For each range of redshifts, we compute the number of pairs in narrow bins of redshift of width 0.1 around a bin center $z_c$. Lastly, we assume a survey area of $10^4$ deg$^2$. We show the resulting percent uncertainties as a function of mass bin center in Fig. ~\ref{fig:mass_unc}, for the lowest and highest redshift bin of each DESI estimate. 

In addition to these results, we include forecasts for Stage-IV stacked weak lensing (WL) given in \cite{WEINBERG201387} (see their Fig. 28). These mass uncertainties are also given for a survey area of $10^4$ deg$^2$ and redshift bins of width $0.1$ for thresholds of halo mass. For a more direct comparison, we estimate the binned WL uncertainties $\sigma_{\rm WL, bin}$ from these 'thresholded' uncertainties $\sigma_{\rm WL, thresh}$ as
\begin{equation}
    \sigma_{\rm WL, bin} = \sqrt{\frac{N_{\rm halos, thresh}}{N_{\rm halos, bin}}}\sigma_{\rm WL, thresh},
\end{equation}
 where $N_{\rm halos, x}$ refers to the simulated number of halos within that mass bin/threshold. Comparing with the uncertainties from the LOS velocity distributions, we forecast accuracy comparable to WL for our highest redshift bins. For our lowest two mass bins and for the highest redshift bins of the BGS and LRG estimates, we forecast \it greater \rm accuracy relative to WL, pushing into the sub-percent level. We also note that our estimates have the greatest accuracy at high redshift, in contrast to WL.   

\section{Anticipated Systematics}\label{sec:syst}

\begin{figure*}
    \centering
    \includegraphics[width=\textwidth]{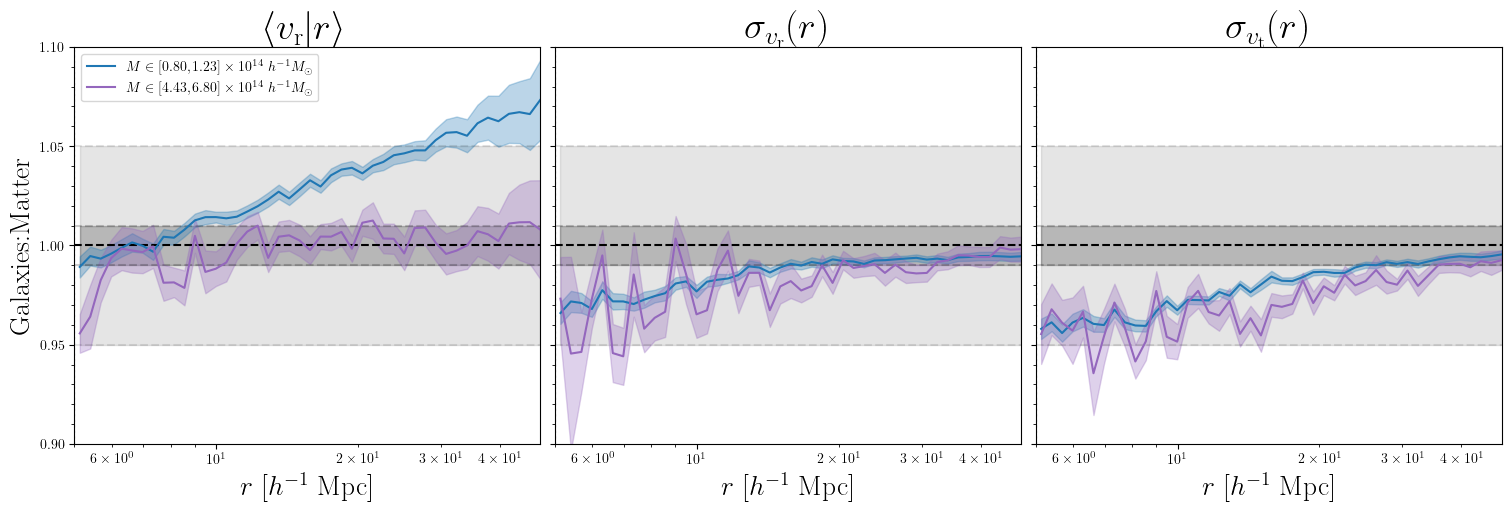}
    \caption{Ratios of the cluster-galaxy mean radial velocity, radial dispersion, and tangential dispersion to those for an analogous cluster-matter selection in MDPL2. The colored lines correspond to measurements within our lowest and next-to-highest cluster mass bins, with error bands coming from jackknifing clusters in sub-boxes of the simulation box. The dark and light gray shaded regions indicate 1\% and 5\% differences, respectively.}
    \label{fig:vel_bias}
\end{figure*}

 While our forecast results in a promising
statistical floor, there are a number of systematic effects that will impact a fully realized mass calibration with the infall region. We discuss these briefly here, but postpone a full characterization of the systematic error budget to a future work.
\vspace{6pt}

\noindent 

\textbf{Simulation measurement uncertainties and dependence on cosmology:} Our forecast does not currently account for uncertainties in our model parameters.  Moreover, the model parameters are taken to be cosmology independent.  Both of these effects can be minimized through a dedicated simulation campaign, see e.g. \citet{2024MNRAS.533.4081R} for an example of this approach within the context of the GIK model.  However, simulations that enable the characterization of the velocity fields of galaxy clusters require large simulations, while the use of \universemachine\ for populating galaxies requires high resolution.  Consequently, the computational demands for such a campaign are non-trivial.

One possible way to reduce the computational footprint is to characterize the velocity bias of galaxies relative to the dark matter, and to study the velocity field of galaxy clusters using lower mass resolution suites of simulations, e.g. \citep{quijote}. Assuming the cluster halos can be taken to be at rest, the equivalence principle suggests that there should be no velocity bias.  We can calculate the small corrections due the halo moving in response to the galaxies using a simple conservation of energy argument: given a mass $M$ and a mass $m$ start at rest, the velocity of mass $m$ is biased with respect to the velocity of point mass of infinitesimal mass via
\begin{equation}
    b_{\rm v}=\left( 1 + \frac{m}{M}\right)^{1/2}.
\end{equation}
Thus, the resulting velocity bias is expected to be scale independent. For the $\sim 10^{10}\ h^{-1} {\rm M}_\odot$ stellar masses we have been using, the bias should fall well below 1\%.

Figure~\ref{fig:vel_bias} compares measured velocity statistics in the simulation.  The left panel shows the mean of the radial velocities of our galaxy catalog to that for an analogous selection of dark matter particles in the MDPL2 simulation, while the central and right panels compare the radial and tangential velocity dispersions of the galaxies and particles.We see a strong, scale-dependent velocity bias signal that is well in excess of that predicted by conservation of energy.  We conjecture this kind of velocity bias may arise from interactions of the galaxy with the cluster's correlated structure.  That is, the observed velocity bias is sourced by a non-zero three-point correlation function. Developing such a bias model is beyond the scope of this work, but doing so is a necessity if the statistical promise of this method is to be realized.

\vspace{6pt}

\noindent 

\textbf{Uncertainties in the galaxy--halo connection and baryonic physics:} The fact that our simulations recover a non-unity velocity bias implies that the relation between galaxy velocity and dark matter particles must depend on the precise galaxy selection criteria and the associated galaxy--halo connection.  This can be achieved, for instance, through a parameterization of the velocity bias observed in Fig.~\ref{fig:vel_bias}. Implementation in data will require demonstrating the robustness of the method on multiple galaxy samples selected from hydrodynamic simulations, e.g. Flamingo \citep{flamingo}.  This will help characterize not only whether the velocity bias prescriptions can adequately capture the uncertainties associated with the galaxy--halo connection, but also whether baryonic feedback can bias the inference process. We leave these tasks to future work.

\vspace{6pt}

\noindent 

\textbf{Observational Systematics:} To measure the cluster velocity probability distributions, one must stack galaxy clusters \citep[e.g.][]{Tomooka}. This can only be done using cluster observables, e.g. richness or SZ-decrement.  The mean and scatter of the observable--mass relation are therefore necessary parameters that need to be marginalized over \citep{miyatake2025cosmologygalaxyclusters}.  More broadly, clusters can also suffer from selection effects, e.g. galaxies along the LOS of a cluster being spuriously assigned cluster membership, thereby impacting the richness while simultaneously biasing the velocity field relative to randomly selected halos. Finally, cluster miscentering is inevitable observational systematic which convolves the infall velocity distributions with the velocity probability distribution of nearby galaxies.  For a cluster catalog such as \texttt{redMaPPer} one expects $\approx 15\%$ of the clusters to be miscentered (\cite{Zhang_2019}). Consequently, it is very likely that a naive application of our model to an obervational cluster sample will result in biased cosmological inference \citep[e.g.][]{desy1_clusters}. In practice, the most robust way to model these effects is through a simulation-based forward modeling inference framework that can adequately account for and marginalize over these observational systematics.  A recent example of such an analysis framework within the context of weak gravitational lensing is that of \citet{salcedo_etal25}.  The results in this paper strongly suggest that adopting this approach to modeling cluster infall velocities would be worthwhile.

\vspace{6pt}

\textbf{Spectroscopic Redshift Errors:} We do not anticipate spectroscopic redshift errors to be a significant source of systematic uncertainty.  Statistical errors in the redshift due to noise in the spectra can be incroporated into the model straightforwardly by convolving the line-of-sight velocity dispersions with the appropriate velocity error.  Catastrophic redshift errors could in principle be problematic, but their incidence is so low \citep[$<1\%$][]{Tomooka,he2025impactspectroscopicredshifterrors} we do not anticipate them to be a significant issue for these type of analysis.

\section{Summary and Conclusions}\label{sec:summary}
We have developed a model for the joint distribution of radial and tangential velocities $P(v_{\rm r}, v_{\rm t}|r, M)$ for scales beyond $r= 5 \ h^{-1}$Mpc. Our model builds on previous work such the the GIK model while taking advantage of observed parameter degeneracies and explicitly parameterizing halo mass dependence. In order to compute the projected LOS velocity distributions $P(v_{\mathrm{LOS}}|R, M)$ that correspond to the joint velocity distributions $P(\vr, \vt|r, M)$, we additionally rely on the model for the halo-galaxy correlation function expounded in \cite{PhysRevD.111.043527}, specifically the infalling component, to determine the necessary weighting along the LOS. We are able to describe these distributions at high accuracy, as is required to utilize them for competitive constraints on cluster masses and cosmology. 

Forecasting the uncertainty in our mass calibration with estimates for the density of DESI tracers, we find that we can obtain sub-percent level accuracy on median cluster mass, competitive with and potentially better than stacked WL constraints. Importantly, this accuracy is achieved with infalling galaxies on the outskirts of clusters, beyond $R=5\ h^{-1}$ Mpc, which we anticipate will be significantly less sensitive to baryonic effects. However, our forecast assumes perfect model calibration and neglects cosmology dependence, which will impact the forecasted mass uncertainty. Additionally, masses recovered from this approach may be subject to bias of the infalling velocities of galaxies with respect to the infalling matter, if any,  and observational systematics such as selection effects and miscentering. Nevertheless, these results provide strong motivation for cluster cosmological analysis in which mass calibration is pursued using the infall regions of galaxy clusters.

\section*{Data Availability}
The MDPL2 halo and UM galaxy catalogues are publicly available and can be found at \url{https://www.peterbehroozi.com/data.html}~\cite{Behroozi_UniverseMachine_MDPL2}.  The CosmoSim database used in this paper is a service by the Leibniz-Institute for Astrophysics Potsdam (AIP). The MultiDark database was developed in cooperation
with the Spanish MultiDark Consolider Project CSD2009-
00064. Other data products underlying the results discussed in this paper will be shared upon reasonable request.

\begin{acknowledgments}
The authors gratefully acknowledge early contributions from Canpo Su, helpful conversations with Andr\'es Salcedo and Tomomi Sunayama, and constructive feedback from the anonymous referee. CS and ER received funding for this
work from the Department of Energy (DOE) grant DE-SC0009913.  CS and ER also gratefully acknowledge funding from the Research Corporation for the Advancement of Science, which funded the early stages of this work. 
\end{acknowledgments}

\appendix

\bibliography{mybib} 

@PREAMBLE{
 "\providecommand{\noopsort}[1]{}" 
 # "\providecommand{\singleletter}[1]{#1}%" 
}

@ARTICLE{lam_etal12,
       author = {{Lam}, Tsz Yan and {Nishimichi}, Takahiro and {Schmidt}, Fabian and {Takada}, Masahiro},
        title = "{Testing Gravity with the Stacked Phase Space around Galaxy Clusters}",
      journal = {\prl},
         year = 2012,
        month = aug,
       volume = {109},
       number = {5},
          eid = {051301},
        pages = {051301},
          doi = {10.1103/PhysRevLett.109.051301},
archivePrefix = {arXiv},
       eprint = {1202.4501},
 primaryClass = {astro-ph.CO},
       adsurl = {https://ui.adsabs.harvard.edu/abs/2012PhRvL.109e1301L}
}

@ARTICLE{desy1_clusters,
       author = {{Abbott}, T.~M.~C. and {Aguena}, M. and {Alarcon}, A. and {Allam}, S. and {Allen}, S. and {Annis}, J. and {Avila}, S. and {Bacon}, D. and {Bechtol}, K. and {Bermeo}, A. and {Bernstein}, G.~M. and {Bertin}, E. and {Bhargava}, S. and {Bocquet}, S. and {Brooks}, D. and {Brout}, D. and {Buckley-Geer}, E. and {Burke}, D.~L. and {Carnero Rosell}, A. and {Carrasco Kind}, M. and {Carretero}, J. and {Castander}, F.~J. and {Cawthon}, R. and {Chang}, C. and {Chen}, X. and {Choi}, A. and {Costanzi}, M. and {Crocce}, M. and {da Costa}, L.~N. and {Davis}, T.~M. and {De Vicente}, J. and {DeRose}, J. and {Desai}, S. and {Diehl}, H.~T. and {Dietrich}, J.~P. and {Dodelson}, S. and {Doel}, P. and {Drlica-Wagner}, A. and {Eckert}, K. and {Eifler}, T.~F. and {Elvin-Poole}, J. and {Estrada}, J. and {Everett}, S. and {Evrard}, A.~E. and {Farahi}, A. and {Ferrero}, I. and {Flaugher}, B. and {Fosalba}, P. and {Frieman}, J. and {Garc{\'\i}a-Bellido}, J. and {Gatti}, M. and {Gaztanaga}, E. and {Gerdes}, D.~W. and {Giannantonio}, T. and {Giles}, P. and {Grandis}, S. and {Gruen}, D. and {Gruendl}, R.~A. and {Gschwend}, J. and {Gutierrez}, G. and {Hartley}, W.~G. and {Hinton}, S.~R. and {Hollowood}, D.~L. and {Honscheid}, K. and {Hoyle}, B. and {Huterer}, D. and {James}, D.~J. and {Jarvis}, M. and {Jeltema}, T. and {Johnson}, M.~W.~G. and {Johnson}, M.~D. and {Kent}, S. and {Krause}, E. and {Kron}, R. and {Kuehn}, K. and {Kuropatkin}, N. and {Lahav}, O. and {Li}, T.~S. and {Lidman}, C. and {Lima}, M. and {Lin}, H. and {MacCrann}, N. and {Maia}, M.~A.~G. and {Mantz}, A. and {Marshall}, J.~L. and {Martini}, P. and {Mayers}, J. and {Melchior}, P. and {Mena-Fern{\'a}ndez}, J. and {Menanteau}, F. and {Miquel}, R. and {Mohr}, J.~J. and {Nichol}, R.~C. and {Nord}, B. and {Ogando}, R.~L.~C. and {Palmese}, A. and {Paz-Chinch{\'o}n}, F. and {Plazas}, A.~A. and {Prat}, J. and {Rau}, M.~M. and {Romer}, A.~K. and {Roodman}, A. and {Rooney}, P. and {Rozo}, E. and {Rykoff}, E.~S. and {Sako}, M. and {Samuroff}, S. and {S{\'a}nchez}, C. and {Sanchez}, E. and {Saro}, A. and {Scarpine}, V. and {Schubnell}, M. and {Scolnic}, D. and {Serrano}, S. and {Sevilla-Noarbe}, I. and {Sheldon}, E. and {Smith}, J. Allyn. and {Smith}, M. and {Suchyta}, E. and {Swanson}, M.~E.~C. and {Tarle}, G. and {Thomas}, D. and {To}, C. and {Troxel}, M.~A. and {Tucker}, D.~L. and {Varga}, T.~N. and {von der Linden}, A. and {Walker}, A.~R. and {Wechsler}, R.~H. and {Weller}, J. and {Wilkinson}, R.~D. and {Wu}, H. and {Yanny}, B. and {Zhang}, Y. and {Zhang}, Z. and {Zuntz}, J. and {DES Collaboration}},
        title = "{Dark Energy Survey Year 1 Results: Cosmological constraints from cluster abundances and weak lensing}",
      journal = {\prd},
         year = 2020,
        month = jul,
       volume = {102},
       number = {2},
          eid = {023509},
        pages = {023509},
          doi = {10.1103/PhysRevD.102.023509},
archivePrefix = {arXiv},
       eprint = {2002.11124},
 primaryClass = {astro-ph.CO},
       adsurl = {https://ui.adsabs.harvard.edu/abs/2020PhRvD.102b3509A}
}

@ARTICLE{flamingo,
       author = {{Helly}, John C. and {McGibbon}, Robert J. and {Schaye}, Joop and {Schaller}, Matthieu and {McDonald}, William and {Braspenning}, Joey and {Broxterman}, Jeger C. and {Costello}, Emily E. and {Elbers}, Willem and {Forouhar Moreno}, Victor J. and {Frenk}, Carlos S. and {Jenkins}, Adrian and {Kugel}, Roi and {McCarthy}, Ian G. and {Salcido}, Jaime and {van Daalen}, Marcel P. and {Vandenbroucke}, Bert and {Yang}, Tianyi},
        title = "{The FLAMINGO simulations data release}",
      journal = {arXiv e-prints},
         year = 2026,
        month = apr,
          eid = {arXiv:2604.24324},
        pages = {arXiv:2604.24324},
          doi = {10.48550/arXiv.2604.24324},
archivePrefix = {arXiv},
       eprint = {2604.24324},
 primaryClass = {astro-ph.CO},
       adsurl = {https://ui.adsabs.harvard.edu/abs/2026arXiv260424324H}
}

@ARTICLE{quijote,
       author = {{Villaescusa-Navarro}, Francisco and {Hahn}, ChangHoon and {Massara}, Elena and {Banerjee}, Arka and {Delgado}, Ana Maria and {Ramanah}, Doogesh Kodi and {Charnock}, Tom and {Giusarma}, Elena and {Li}, Yin and {Allys}, Erwan and {Brochard}, Antoine and {Uhlemann}, Cora and {Chiang}, Chi-Ting and {He}, Siyu and {Pisani}, Alice and {Obuljen}, Andrej and {Feng}, Yu and {Castorina}, Emanuele and {Contardo}, Gabriella and {Kreisch}, Christina D. and {Nicola}, Andrina and {Alsing}, Justin and {Scoccimarro}, Roman and {Verde}, Licia and {Viel}, Matteo and {Ho}, Shirley and {Mallat}, Stephane and {Wandelt}, Benjamin and {Spergel}, David N.},
        title = "{The Quijote Simulations}",
      journal = {\apjs},
         year = 2020,
        month = sep,
       volume = {250},
       number = {1},
          eid = {2},
        pages = {2},
          doi = {10.3847/1538-4365/ab9d82},
archivePrefix = {arXiv},
       eprint = {1909.05273},
 primaryClass = {astro-ph.CO},
       adsurl = {https://ui.adsabs.harvard.edu/abs/2020ApJS..250....2V}
}

@ARTICLE{zu_etal14,
       author = {{Zu}, Ying and {Weinberg}, David H. and {Jennings}, Elise and {Li}, Baojiu and {Wyman}, Mark},
        title = "{Galaxy infall kinematics as a test of modified gravity}",
      journal = {\mnras},
         year = 2014,
        month = dec,
       volume = {445},
       number = {2},
        pages = {1885-1897},
          doi = {10.1093/mnras/stu1739},
archivePrefix = {arXiv},
       eprint = {1310.6768},
 primaryClass = {astro-ph.CO},
       adsurl = {https://ui.adsabs.harvard.edu/abs/2014MNRAS.445.1885Z}
}

@ARTICLE{Tomooka,
       author = {{Tomooka}, Paxton and {Rozo}, Eduardo and {Wagoner}, Erika L. and {Aung}, Han and {Nagai}, Daisuke and {Safonova}, Sasha},
        title = "{Clusters have edges: the projected phase-space structure of SDSS redMaPPer clusters}",
      journal = {\mnras},
         year = 2020,
        month = nov,
       volume = {499},
       number = {1},
        pages = {1291-1299},
          doi = {10.1093/mnras/staa2841},
archivePrefix = {arXiv},
       eprint = {2003.11555},
 primaryClass = {astro-ph.CO},
       adsurl = {https://ui.adsabs.harvard.edu/abs/2020MNRAS.499.1291T}
}

@ARTICLE{2016MNRAS.457.4340K,
       author = {{Klypin}, Anatoly and {Yepes}, Gustavo and {Gottl{\"o}ber}, Stefan and {Prada}, Francisco and {He{\ss}}, Steffen},
        title = "{MultiDark simulations: the story of dark matter halo concentrations and density profiles}",
      journal = {\mnras},
         year = 2016,
        month = apr,
       volume = {457},
       number = {4},
        pages = {4340-4359},
          doi = {10.1093/mnras/stw248},
archivePrefix = {arXiv},
       eprint = {1411.4001},
 primaryClass = {astro-ph.CO},
       adsurl = {https://ui.adsabs.harvard.edu/abs/2016MNRAS.457.4340K}
}

@ARTICLE{redmapper,
       author = {{Rykoff}, E.~S. and {Rozo}, E. and {Busha}, M.~T. and {Cunha}, C.~E. and {Finoguenov}, A. and {Evrard}, A. and {Hao}, J. and {Koester}, B.~P. and {Leauthaud}, A. and {Nord}, B. and {Pierre}, M. and {Reddick}, R. and {Sadibekova}, T. and {Sheldon}, E.~S. and {Wechsler}, R.~H.},
        title = "{redMaPPer. I. Algorithm and SDSS DR8 Catalog}",
      journal = {\apj},
         year = 2014,
        month = apr,
       volume = {785},
       number = {2},
          eid = {104},
        pages = {104},
          doi = {10.1088/0004-637X/785/2/104},
archivePrefix = {arXiv},
       eprint = {1303.3562},
 primaryClass = {astro-ph.CO},
       adsurl = {https://ui.adsabs.harvard.edu/abs/2014ApJ...785..104R}
}

@ARTICLE{redmapper2,
       author = {{Rykoff}, E.~S. and {Rozo}, E. and {Hollowood}, D. and {Bermeo-Hernandez}, A. and {Jeltema}, T. and {Mayers}, J. and {Romer}, A.~K. and {Rooney}, P. and {Saro}, A. and {Vergara Cervantes}, C. and {Wechsler}, R.~H. and {Wilcox}, H. and {Abbott}, T.~M.~C. and {Abdalla}, F.~B. and {Allam}, S. and {Annis}, J. and {Benoit-L{\'e}vy}, A. and {Bernstein}, G.~M. and {Bertin}, E. and {Brooks}, D. and {Burke}, D.~L. and {Capozzi}, D. and {Carnero Rosell}, A. and {Carrasco Kind}, M. and {Castander}, F.~J. and {Childress}, M. and {Collins}, C.~A. and {Cunha}, C.~E. and {D'Andrea}, C.~B. and {da Costa}, L.~N. and {Davis}, T.~M. and {Desai}, S. and {Diehl}, H.~T. and {Dietrich}, J.~P. and {Doel}, P. and {Evrard}, A.~E. and {Finley}, D.~A. and {Flaugher}, B. and {Fosalba}, P. and {Frieman}, J. and {Glazebrook}, K. and {Goldstein}, D.~A. and {Gruen}, D. and {Gruendl}, R.~A. and {Gutierrez}, G. and {Hilton}, M. and {Honscheid}, K. and {Hoyle}, B. and {James}, D.~J. and {Kay}, S.~T. and {Kuehn}, K. and {Kuropatkin}, N. and {Lahav}, O. and {Lewis}, G.~F. and {Lidman}, C. and {Lima}, M. and {Maia}, M.~A.~G. and {Mann}, R.~G. and {Marshall}, J.~L. and {Martini}, P. and {Melchior}, P. and {Miller}, C.~J. and {Miquel}, R. and {Mohr}, J.~J. and {Nichol}, R.~C. and {Nord}, B. and {Ogando}, R. and {Plazas}, A.~A. and {Reil}, K. and {Sahl{\'e}n}, M. and {Sanchez}, E. and {Santiago}, B. and {Scarpine}, V. and {Schubnell}, M. and {Sevilla-Noarbe}, I. and {Smith}, R.~C. and {Soares-Santos}, M. and {Sobreira}, F. and {Stott}, J.~P. and {Suchyta}, E. and {Swanson}, M.~E.~C. and {Tarle}, G. and {Thomas}, D. and {Tucker}, D. and {Uddin}, S. and {Viana}, P.~T.~P. and {Vikram}, V. and {Walker}, A.~R. and {Zhang}, Y. and {DES Collaboration}},
        title = "{The RedMaPPer Galaxy Cluster Catalog From DES Science Verification Data}",
      journal = {\apjs},
         year = 2016,
        month = may,
       volume = {224},
       number = {1},
          eid = {1},
        pages = {1},
          doi = {10.3847/0067-0049/224/1/1},
archivePrefix = {arXiv},
       eprint = {1601.00621},
 primaryClass = {astro-ph.CO},
       adsurl = {https://ui.adsabs.harvard.edu/abs/2016ApJS..224....1R}
}

@ARTICLE{2013ApJ...762..109B,
       author = {{Behroozi}, Peter S. and {Wechsler}, Risa H. and {Wu}, Hao-Yi},
        title = "{The ROCKSTAR Phase-space Temporal Halo Finder and the Velocity Offsets of Cluster Cores}",
      journal = {\apj},
         year = 2013,
        month = jan,
       volume = {762},
       number = {2},
          eid = {109},
        pages = {109},
          doi = {10.1088/0004-637X/762/2/109},
archivePrefix = {arXiv},
       eprint = {1110.4372},
 primaryClass = {astro-ph.CO},
       adsurl = {https://ui.adsabs.harvard.edu/abs/2013ApJ...762..109B}
}

@ARTICLE{2013ApJ...763...18B,
       author = {{Behroozi}, Peter S. and {Wechsler}, Risa H. and {Wu}, Hao-Yi and {Busha}, Michael T. and {Klypin}, Anatoly A. and {Primack}, Joel R.},
        title = "{Gravitationally Consistent Halo Catalogs and Merger Trees for Precision Cosmology}",
      journal = {\apj},
         year = 2013,
        month = jan,
       volume = {763},
       number = {1},
          eid = {18},
        pages = {18},
          doi = {10.1088/0004-637X/763/1/18},
archivePrefix = {arXiv},
       eprint = {1110.4370},
 primaryClass = {astro-ph.CO},
       adsurl = {https://ui.adsabs.harvard.edu/abs/2013ApJ...763...18B}
}

@ARTICLE{2019MNRAS.488.3143B,
       author = {{Behroozi}, Peter and {Wechsler}, Risa H. and {Hearin}, Andrew P. and {Conroy}, Charlie},
        title = "{UNIVERSEMACHINE: The correlation between galaxy growth and dark matter halo assembly from z = 0-10}",
      journal = {\mnras},
         year = 2019,
        month = sep,
       volume = {488},
       number = {3},
        pages = {3143-3194},
          doi = {10.1093/mnras/stz1182},
archivePrefix = {arXiv},
       eprint = {1806.07893},
 primaryClass = {astro-ph.GA},
       adsurl = {https://ui.adsabs.harvard.edu/abs/2019MNRAS.488.3143B}
}

@ARTICLE{zuweinberg13,
       author = {{Zu}, Ying and {Weinberg}, David H.},
        title = "{The redshift-space cluster-galaxy cross-correlation function - I. Modelling galaxy infall on to Millennium simulation clusters and SDSS groups}",
      journal = {\mnras},
         year = 2013,
        month = jun,
       volume = {431},
       number = {4},
        pages = {3319-3337},
          doi = {10.1093/mnras/stt411},
archivePrefix = {arXiv},
       eprint = {1211.1379},
 primaryClass = {astro-ph.CO},
       adsurl = {https://ui.adsabs.harvard.edu/abs/2013MNRAS.431.3319Z}
}

@article{10.1093/mnras/staa3994,
    author = {Aung, Han and Nagai, Daisuke and Rozo, Eduardo and García, Rafael},
    title = {The phase-space structure of dark matter haloes},
    journal = {Monthly Notices of the Royal Astronomical Society},
    volume = {502},
    number = {1},
    pages = {1041-1047},
    year = {2020},
    month = {12},
    issn = {0035-8711},
    doi = {10.1093/mnras/staa3994},
    url = {https://doi.org/10.1093/mnras/staa3994},
    eprint = {https://academic.oup.com/mnras/article-pdf/502/1/1041/36169658/staa3994.pdf},
}

@article{PhysRevD.111.043527,
  title = {Dynamics-based halo model for large scale structure},
  author = {Salazar, Edgar M. and Rozo, Eduardo and Garc\'{\i}a, Rafael and Kokron, Nickolas and Adhikari, Susmita and Diemer, Benedikt and Osinga, Calvin},
  journal = {Phys. Rev. D},
  volume = {111},
  issue = {4},
  pages = {043527},
  numpages = {17},
  year = {2025},
  month = {Feb},
  publisher = {American Physical Society},
  doi = {10.1103/PhysRevD.111.043527},
  url = {https://link.aps.org/doi/10.1103/PhysRevD.111.043527}
}

@article{10.1093/mnras/stad601,
    author = {Aung, Han and Nagai, Daisuke and Rozo, Eduardo and Wolfe, Brandon and Adhikari, Susmita},
    title = {Accurate model of the projected velocity distribution of galaxies in dark matter haloes},
    journal = {Monthly Notices of the Royal Astronomical Society},
    volume = {521},
    number = {3},
    pages = {3981-3990},
    year = {2023},
    month = {03},
    issn = {0035-8711},
    doi = {10.1093/mnras/stad601},
    url = {https://doi.org/10.1093/mnras/stad601},
    eprint = {https://academic.oup.com/mnras/article-pdf/521/3/3981/49672134/stad601.pdf},
}

@ARTICLE{2024MNRAS.533.4081R,
       author = {{Robertson}, Andrew and {Huff}, Eric and {Markovi{\v{c}}}, Katarina and {Li}, Baojiu},
        title = "{Modelling the redshift-space cluster-galaxy correlation function on Mpc scales with emulation of the pairwise velocity distribution}",
      journal = {\mnras},
         year = 2024,
        month = oct,
       volume = {533},
       number = {4},
        pages = {4081-4103},
          doi = {10.1093/mnras/stae1980},
archivePrefix = {arXiv},
       eprint = {2406.01527},
 primaryClass = {astro-ph.CO},
       adsurl = {https://ui.adsabs.harvard.edu/abs/2024MNRAS.533.4081R}
}

@ARTICLE{10.1093/mnras/stz2227,
    author = {Hamabata, Akinari and Oguri, Masamune and Nishimichi, Takahiro},
    title = {Constraining cluster masses from the stacked phase space distribution at large radii},
    journal = {Monthly Notices of the Royal Astronomical Society},
    volume = {489},
    number = {1},
    pages = {1344-1356},
    year = {2019},
    month = {08},
    issn = {0035-8711},
    doi = {10.1093/mnras/stz2227},
    url = {https://doi.org/10.1093/mnras/stz2227},
    eprint = {https://academic.oup.com/mnras/article-pdf/489/1/1344/29229772/stz2227.pdf},
}

@article{PhysRevD.70.083007,
  title = {Redshift-space distortions, pairwise velocities, and nonlinearities},
  author = {Scoccimarro, Rom\'an},
  journal = {Phys. Rev. D},
  volume = {70},
  issue = {8},
  pages = {083007},
  numpages = {19},
  year = {2004},
  month = {Oct},
  publisher = {American Physical Society},
  doi = {10.1103/PhysRevD.70.083007},
  url = {https://link.aps.org/doi/10.1103/PhysRevD.70.083007}
}

@misc{he2025impactspectroscopicredshifterrors,
      title={The Impact of Spectroscopic Redshift Errors on Cosmological Measurements}, 
      author={Shengyu He and Jiaxi Yu and Antoine Rocher and Daniel Forero-Sánchez and Jean-Paul Kneib and Cheng Zhao and Etienne Burtin and Jiamin Hou},
      year={2025},
      eprint={2508.21182},
      archivePrefix={arXiv},
      primaryClass={astro-ph.CO},
      url={https://arxiv.org/abs/2508.21182}, 
}

@ARTICLE{salcedo_etal25,
       author = {{Salcedo}, Andr{\'e}s N. and {Rozo}, Eduardo and {Wu}, Hao-Yi and {Weinberg}, David H. and {Chiploonkar}, Pranav and {To}, Chun-Hao and {Cao}, Shulei and {Rykoff}, Eli S. and {Marcelina Gountanis}, Nicole and {Zhou}, Conghao},
        title = "{Cosmological Constraints from Dark Energy Survey Year 1 Cluster Lensing and Abundances with Simulation-based Forward-Modeling}",
      journal = {arXiv e-prints},
         year = 2025,
        month = oct,
          eid = {arXiv:2510.25706},
        pages = {arXiv:2510.25706},
          doi = {10.48550/arXiv.2510.25706},
archivePrefix = {arXiv},
       eprint = {2510.25706},
 primaryClass = {astro-ph.CO},
       adsurl = {https://ui.adsabs.harvard.edu/abs/2025arXiv251025706S}
}

@misc{miyatake2025cosmologygalaxyclusters,
      title={Cosmology with Galaxy Clusters}, 
      author={Hironao Miyatake},
      year={2025},
      eprint={2505.07697},
      archivePrefix={arXiv},
      primaryClass={astro-ph.CO},
      url={https://arxiv.org/abs/2505.07697}, 
}

@article{WEINBERG201387,
title = {Observational probes of cosmic acceleration},
journal = {Physics Reports},
volume = {530},
number = {2},
pages = {87-255},
year = {2013},
note = {Observational Probes of Cosmic Acceleration},
issn = {0370-1573},
doi = {https://doi.org/10.1016/j.physrep.2013.05.001},
url = {https://www.sciencedirect.com/science/article/pii/S0370157313001592},
author = {David H. Weinberg and Michael J. Mortonson and Daniel J. Eisenstein and Christopher Hirata and Adam G. Riess and Eduardo Rozo}
}

@INPROCEEDINGS{1995AAS...187.9504B,
       author = {{Bryan}, G.~L. and {Norman}, M.~L.},
        title = "{Simulating X-ray Clusters with Adaptive Mesh Refinement}",
    booktitle = {American Astronomical Society Meeting Abstracts},
         year = 1995,
       series = {American Astronomical Society Meeting Abstracts},
       volume = {187},
        month = dec,
          eid = {95.04},
        pages = {95.04},
       adsurl = {https://ui.adsabs.harvard.edu/abs/1995AAS...187.9504B}
}

@software{emcee,
       author = {{Foreman-Mackey}, Daniel and {Conley}, Alex and {Meierjurgen Farr}, Will and {Hogg}, David W. and {Lang}, Dustin and {Marshall}, Phil and {Price-Whelan}, Adrian and {Sanders}, Jeremy and {Zuntz}, Joe},
        title = "{emcee: The MCMC Hammer}",
 howpublished = {Astrophysics Source Code Library, record ascl:1303.002},
         year = 2013,
        month = mar,
          eid = {ascl:1303.002},
archivePrefix = {ascl},
       eprint = {1303.002},
       adsurl = {https://ui.adsabs.harvard.edu/abs/2013ascl.soft03002F}
}

@article{tr6y-kpc6,
  title = {DESI DR2 results. II. Measurements of baryon acoustic oscillations and cosmological constraints},
  author = {Abdul Karim, M. and Aguilar, J. and Ahlen, S. and Alam, S. and Allen, L. and Prieto, C. Allende and Alves, O. and Anand, A. and Andrade, U. and Armengaud, E. and Aviles, A. and Bailey, S. and Baltay, C. and Bansal, P. and Bault, A. and Behera, J. and BenZvi, S. and Bianchi, D. and Blake, C. and Brieden, S. and Brodzeller, A. and Brooks, D. and Buckley-Geer, E. and Burtin, E. and Calderon, R. and Canning, R. and Rosell, A. Carnero and Carrilho, P. and Casas, L. and Castander, F. J. and Charles, M. and Chaussidon, E. and Chaves-Montero, J. and Chebat, D. and Chen, X. and Claybaugh, T. and Cole, S. and Cooper, A. P. and Cuceu, A. and Dawson, K. S. and de la Macorra, A. and de Mattia, A. and Deiosso, N. and Della Costa, J. and Demina, R. and Dey, A. and Dey, B. and Ding, Z. and Doel, P. and Edelstein, J. and Eisenstein, D. J. and Elbers, W. and Fagrelius, P. and Fanning, K. and Fern\'andez-Garc\'{\i}a, E. and Ferraro, S. and Font-Ribera, A. and Forero-Romero, J. E. and Frenk, C. S. and Garcia-Quintero, C. and Garrison, L. H. and Gazta\~naga, E. and Gil-Mar\'{\i}n, H. and Gontcho, S. Gontcho A. and Gonzalez, D. and Gonzalez-Morales, A. X. and Gordon, C. and Green, D. and Gutierrez, G. and Guy, J. and Hadzhiyska, B. and Hahn, C. and He, S. and Herbold, M. and Herrera-Alcantar, H. K. and Ho, M.-F. and Honscheid, K. and Howlett, C. and Huterer, D. and Ishak, M. and Juneau, S. and Kamble, N. V. and Kara\ifmmode \mbox{\c{c}}\else \c{c}\fi{}ayl, N. G. and Kehoe, R. and Kent, S. and Kim, A. G. and Kirkby, D. and Kisner, T. and Koposov, S. E. and Kremin, A. and Krolewski, A. and Lahav, O. and Lamman, C. and Landriau, M. and Lang, D. and Lasker, J. and Le Goff, J. M. and Le Guillou, L. and Leauthaud, A. and Levi, M. E. and Li, Q. and Li, T. S. and Lodha, K. and Lokken, M. and Lozano-Rodr\'{\i}guez, F. and Magneville, C. and Manera, M. and Martini, P. and Matthewson, W. L. and Meisner, A. and Mena-Fern\'andez, J. and Menegas, A. and Mergulh\~ao, T. and Miquel, R. and Moustakas, J. and Mu\~noz-Guti\'errez, A. and Mu\~noz-Santos, D. and Myers, A. D. and Nadathur, S. and Naidoo, K. and Napolitano, L. and Newman, J. A. and Niz, G. and Noriega, H. E. and Paillas, E. and Palanque-Delabrouille, N. and Pan, J. and Peacock, J. A. and Ibanez, M. P. and Percival, W. J. and P\'erez-Fern\'andez, A. and P\'erez-R\`afols, I. and Pieri, M. M. and Poppett, C. and Prada, F. and Rabinowitz, D. and Raichoor, A. and Ram\'{\i}rez-P\'erez, C. and Rashkovetskyi, M. and Ravoux, C. and Rich, J. and Rocher, A. and Rockosi, C. and Rohlf, J. and Rom\'an-Herrera, J. O. and Ross, A. J. and Rossi, G. and Ruggeri, R. and Ruhlmann-Kleider, V. and Samushia, L. and Sanchez, E. and Sanders, N. and Schlegel, D. and Schubnell, M. and Seo, H. and Shafieloo, A. and Sharples, R. and Silber, J. and Sinigaglia, F. and Sprayberry, D. and Tan, T. and Tarl\'e, G. and Taylor, P. and Turner, W. and Ure\~na-L\'opez, L. A. and Vaisakh, R. and Valdes, F. and Valogiannis, G. and Vargas-Maga\~na, M. and Verde, L. and Walther, M. and Weaver, B. A. and Weinberg, D. H. and White, M. and Wolfson, M. and Y\`eche, C. and Yu, J. and Zaborowski, E. A. and Zarrouk, P. and Zhai, Z. and Zhang, H. and Zhao, C. and Zhao, G. B. and Zhou, R. and Zou, H.},
  collaboration = {DESI Collaboration},
  journal = {Phys. Rev. D},
  volume = {112},
  issue = {8},
  pages = {083515},
  numpages = {40},
  year = {2025},
  month = {Oct},
  publisher = {American Physical Society},
  doi = {10.1103/tr6y-kpc6},
  url = {https://link.aps.org/doi/10.1103/tr6y-kpc6}
}

@article{10.1093/mnras/staa2249,
    author = {Cuesta-Lazaro, Carolina and Li, Baojiu and Eggemeier, Alexander and Zarrouk, Pauline and Baugh, Carlton M and Nishimichi, Takahiro and Takada, Masahiro},
    title = {Towards a non-Gaussian model of redshift space distortions},
    journal = {Monthly Notices of the Royal Astronomical Society},
    volume = {498},
    number = {1},
    pages = {1175-1193},
    year = {2020},
    month = {08},
    issn = {0035-8711},
    doi = {10.1093/mnras/staa2249},
    url = {https://doi.org/10.1093/mnras/staa2249},
    eprint = {https://academic.oup.com/mnras/article-pdf/498/1/1175/33731014/staa2249.pdf},
}

@article{Zhang_2019,
   title={Dark Energy Surveyed Year 1 results: calibration of cluster mis-centring in the redMaPPer catalogues},
   volume={487},
   ISSN={1365-2966},
   url={http://dx.doi.org/10.1093/mnras/stz1361},
   DOI={10.1093/mnras/stz1361},
   number={2},
   journal={Monthly Notices of the Royal Astronomical Society},
   publisher={Oxford University Press (OUP)},
   author={Zhang, Y and Jeltema, T and Hollowood, D L and Everett, S and Rozo, E and Farahi, A and Bermeo, A and Bhargava, S and Giles, P and Romer, A K and Wilkinson, R and Rykoff, E S and Mantz, A and Diehl, H T and Evrard, A E and Stern, C and Gruen, D and von der Linden, A and Splettstoesser, M and Chen, X and Costanzi, M and Allen, S and Collins, C and Hilton, M and Klein, M and Mann, R G and Manolopoulou, M and Morris, G and Mayers, J and Sahlen, M and Stott, J and Vergara Cervantes, C and Viana, P T P and Wechsler, R H and Allam, S and Avila, S and Bechtol, K and Bertin, E and Brooks, D and Burke, D L and Carnero Rosell, A and Carrasco Kind, M and Carretero, J and Castander, F J and da Costa, L N and De Vicente, J and Desai, S and Dietrich, J P and Doel, P and Flaugher, B and Fosalba, P and Frieman, J and García-Bellido, J and Gaztanaga, E and Gruendl, R A and Gschwend, J and Gutierrez, G and Hartley, W G and Honscheid, K and Hoyle, B and Krause, E and Kuehn, K and Kuropatkin, N and Lima, M and Maia, M A G and Marshall, J L and Melchior, P and Menanteau, F and Miller, C J and Miquel, R and Ogando, R L C and Plazas, A A and Sanchez, E and Scarpine, V and Schindler, R and Serrano, S and Sevilla-Noarbe, I and Smith, M and Soares-Santos, M and Suchyta, E and Swanson, M E C and Tarle, G and Thomas, D and Tucker, D L and Vikram, V and Wester, W and },
   year={2019},
   month=May, pages={2578–2593} }

@misc{Behroozi_UniverseMachine_MDPL2,
  author = {Behroozi, Peter},
  title  = {{UniverseMachine and MDPL2 Data Release}},
  url    = {https://www.peterbehroozi.com/data.html},
  note   = {Accessed: 2026-07-10}
}

\end{document}